\documentclass[pre,twocolumn,twoside,showpacs,superscriptaddress,tightenlines,nofootinbib]{
revtex4-1}
\usepackage{amssymb,graphicx,amsmath,epsf,bm} 
\epsfclipon
\usepackage{color}

\definecolor{nred}{RGB}{224,0,0}
\definecolor{nblue}  {RGB}{28,130,185}
\definecolor{dgreen} {RGB}{78,138,21}
\definecolor{norange}{RGB}{230,120,20}

\newcommand{\be}{\begin{equation}}
\newcommand{\ee}{\end{equation}}

\begin{document}

\title{Precision calculation of critical exponents in the $O(N)$
  universality classes with the nonperturbative renormalization group}

\author{Gonzalo De Polsi}
\affiliation{Instituto de F\'isica, Facultad de Ciencias, Universidad de la
		Rep\'ublica, Igu\'a 4225, 11400, Montevideo, Uruguay}

\author{Ivan Balog}
\affiliation{Institute of Physics, Bijeni\v{c}ka cesta 46, HR-10001 Zagreb, 
Croatia}

\author{Matthieu Tissier}
\affiliation{Sorbonne Universit\'e, CNRS, Laboratoire de Physique Th\'eorique de 
la Mati\`ere Condens\'ee, LPTMC, F-75005 Paris, France}

\author{Nicol\'as Wschebor}
\affiliation{Instituto de F\'isica, Facultad de Ingenier\'ia, Universidad de la
		Rep\'ublica, J.H.y Reissig 565, 11000 Montevideo, Uruguay}

\date{\today}

\begin{abstract}
  We compute the critical exponents $\nu$, $\eta$ and $\omega$ of
  $O(N)$ models for various values of $N$ by implementing the
  derivative expansion of the nonperturbative renormalization group up
  to next-to-next-to-leading order [usually denoted
  $\mathcal{O}(\partial^4)$].  We analyze the behavior of
  this approximation scheme at successive orders and observe an
  apparent convergence with a small parameter -- typically between $1/9$
  and $1/4$ -- compatible with previous studies in the Ising case. This
  allows us to give well-grounded error bars. We obtain a
  determination of critical exponents with a precision which is similar or
  better than those obtained by most field theoretical techniques. We
  also reach a better precision than Monte-Carlo simulations in some
  physically relevant situations. In the $O(2)$ case, where there is a
  longstanding controversy between Monte-Carlo estimates and
  experiments for the specific heat exponent $\alpha$, our results are
  compatible with those of Monte-Carlo but clearly exclude
  experimental values.
\end{abstract}

\maketitle

\section{Introduction}


Systems where microscopic degrees of freedom are stongly coupled are
notoriously difficult to analyze theoretically. This difficulty
becomes even more involved if the system is near a critical point
because of the large number of interacting degrees of freedom that
must be treated simultaneously. From the theoretical viewpoint, two
methods are widely used to study these physical situations. The first
one was introduced by Wilson: the Renormalization Group (RG)
\cite{Wilson:1973jj}. This technique, when used in conjunction with
perturbation theory is able to describe systems with many interacting
degrees of freedom with a small or moderate effective coupling among
infrared degrees of freedom. The perturbative implementation of the RG
\cite{Guida:1998bx,Pelissetto02} has become a fantastic method both in
statistical physics and in quantum field theory when very different
scales are present \cite{ZinnJustin:2002ru}. In the realm of
statistical physics, it has been used to describe both equilibrium and
out-of-equilibrium situations, it can deal with quenched disorder,
long range interactions, etc. A main limitation of this approach is
that it is based on an expansion in some small coupling and it cannot
be applied to systems where no such small parameter is
known. Moreover, the algebraic complexity of the calculation strongly
increases with the order of the expansion. Due to this complexity,
only recently progress have been done \cite{Schnetz:2016fhy} and the
perturbative series have been pushed to 7 loops. Another limitation of
perturbative RG is that the series do not converge in general and one
has to resort to some resummation techniques in order to make precise
predictions. These techniques always involve some unknown parameters
that must be fixed by using some extra criterion, such as the
principle of minimal sensitivity or the principle of fastest apparent
convergence, see below.


The other popular theoretical approach to critical systems is computer
simulations \cite{landau2014guide}. A major asset of these
techniques is their versatility: they can be applied to a large number
of situations -- at criticality or away from criticality -- even when
perturbative RG treatment might be very difficult. At a quantitative
level, high precision estimates of the critical exponents were
obtained by these methods, see \cite{Pelissetto02} for ar review. A
major drawback is that it can require extremely large amounts of
computer time and statistical and systematic errors only decreases
slowly with the size of the simulation. To give an example, for the
Monte-Carlo studies of criticality of the pure Ising model, which are
considered to be the most favorable case numerically, the most
extensive numerical study \cite{Hasenbusch10} reaches lattice sizes of
$L=300$ in 3d, for which 30 years of CPU time are needed. In the case
of the most recent simulation on the XY model
\cite{Hasenbusch:2019jkj}, on which we comment later on, the numerical
load is approximately four times bigger.

There also exist methods which apply only to some particular physical
situations. Among these, let us cite the large-$N$ expansion, high-
and low-temperature expansions. The other method of choice for
studying critical exponents us conformal field
theory \cite{Belavin:1984vu,DiFrancesco:1997nk} which can be applied to
a variety of systems at equilibrium in their critical regime, which
present, on top of scale invariance the whole conformal group. These
methods were first developed in the bidimensional case but were more
recently applied to higher dimensions, through the Conformal Bootstrap
(CB) program \cite{ElShowk:2012ht,El-Showk:2014dwa,Kos:2014bka}. This
led in the recent past to an unprecedented precision on critical
exponents for the Ising model. Such methods are however unable to
access other quantities of physical interest, such as a phase diagram.

The two versatile methods mentioned above -- perturbative
renormalization-group and lattice simulations -- have both their
limitations. In order to overcome some of these, a third, flexible,
method was developed in the 90's
\cite{Wetterich:1992yh,Ellwanger:1993kk,Morris:1993qb,Berges:2000ew,
  Delamotte:2007pf}. It is known as the nonperturbative RG (NPRG
hereafter) or ``functional RG'' or, even ``exact RG''. It is nowadays
widely used in particle physics, solid state physics, statistical
mechanics in and out of equilibrium, quantum gravity, etc. We shall
describe the NPRG in more details below but, in a nutshell, it is
based on an exact RG equation which describes the evolution of an
effective average action when more and more short-distance
fluctuations are integrated over. This equation is too complex to be
solved exactly. In actual calculations, the strategy consists in
looking for approximate solution to this exact equation: instead of
considering the full functional dependence of the effective average
action, one retains only a subset of coupling constants and looks for
(approximate) solutions within this subset. The most popular
approximation scheme is the Derivative Expansion (DE). It consists in
classifying the terms appearing in the effective average action
according to the number of gradients they contain and retaining only
those with up to $s$ gradients. We refer to this approximation as DE
at order $\mathcal O(\partial^s)$ and the leading approximation, where
all derivatives of the field are depreciated, except for a
unrenormalized gradient $(\partial\phi)^2$ is called the Local
Potential Approximation (LPA). This is equivalent to saying that the
$n$-point vertex functions are expanded in powers of the momenta, up
to order $p^s$. Such an expansion is justified if one is interested in
the long-distance properties of the system, see below for more detail.

What is remarkable about the NPRG is that it is very resilient: even
quite crude truncations can lead to qualitatively correct
physics. Until recently, a major drawback of the method was that only
limited knowledge was available concerning the convergence of the
results when richer truncations were considered. This situation
changed last year \cite{Balog:2019rrg}, when it was shown, in the case
of the Ising model, that the results in the DE should converge with a
convergence parameter in-between 1/4 and 1/9, which is indeed not too
large. This theoretical prediction was checked explicitly by computing
critical exponents $\eta$ and $\nu$ in the derivative expansion pushed
up to $\mathcal O(\partial ^6)$. The output of this study is the
determination of the critical exponents $\nu = 0.6300(2)$ and
$\eta = 0.0358(6)$. Remarkably, these are in excellent agreement with
CB values $\nu = 0.629971(4)$ and $\eta = 0.0362978(20)$
\cite{Kos:2014bka}, and better than perturbative 6-loop ones
\cite{Guida:1998bx}. This is an important breakthrough which shows
that the NPRG can be used to obtain precise determinations of critical
exponents, with well-grounded error bars. Note that critical exponents
are just one example of physical quantities that can be computed by
NPRG methods. They provide a good benchmarks to test the
convergence of DE because
other methods, such as CB  and MC, yield very precise determinations of
these quantities. NPRG, however, can be used to determine other
physical quantities, such as a critical temperature, scaling functions and
we expect that the convergence of the DE is governed by the same small
parameter, of the order of 1/4 to 1/9.

Our aim in this paper is twofold. We first show that the convergence
of the DE in O($N$) models is similar to what was found in the Ising
case, with a convergence parameter in-between 1/4 and 1/9. This is
checked explicitly by looking at the convergence of the DE expansion
up to $\mathcal O(\partial^4)$ for the critical exponents $\eta$ and
$\nu$. We also treat the correction to scaling exponent $\omega$ which
was not considered in \cite{Balog:2019rrg}.\footnote{The authors of
  \cite{Balog:2019rrg} had studied this exponent for $N=1$ but did not
  publish it. We acknowledge discussion with them on this topic.} This
enables us to determine convincing error bars. We describe in detail
the methodology used to determine error bars because it is quite
generic and could be used in many applications of the NPRG. The second
aim of this article is to determine the critical exponents for
different values of $N$. This is not only an academic issue because
the O($N$) universality classes for $N$=1, 2, 3 and 4 have direct
physical realizations \cite{ZinnJustin:2002ru}.  Additionally the
limits $N\to 0$ and $N\to -2$ are also of physical interest, being
related to self avoiding random walks \cite{degennes72} and loop
erased random walks \cite{Wiese:2018dow,Wiese:2019xmu} respectively.

The $N=2$ case is of particular interest because it describes the
normal to superfluid transition in Helium-4. Experimental methods led
to a determination of the critical exponent which governs the
singularity of the specific heat with unprecedented precision. By
using hyperscaling relation, this yields an exponent $\nu=0.6709(1)$
\cite{He4exp}. The main limitation on experiments was the variation of
the density of the fluid within the sample caused by gravity and it
was necessary to perform the experiment in the Space Shuttle in order
to obtain a sufficiently homogeneous system.

This high precision experiment triggered an important theoretical
effort to obtain a determination of critical exponents with a similar
precision.  What is curious is that there exists a discrepancy between
the experimental and the most precise Monte-Carlo results
\cite{Hasenbusch:2019jkj}, which reports $\nu=0.67169(7)$. These two
results are not compatible. Other field-theoretical results based on
perturbative renormalization-group led to too large error bars to
settle the controversy. One of the main results of this article is the
determination of the critical exponent $\nu=0.6716(6)$ which is
compatible with Monte-Carlo results, but not with experimental
ones. We should mention that, during the completion of this article, a
theoretical result based on CB was reported
\cite{chester2019carving}, which leads to the same conclusion, see
below for more detail.

The article is organized as follows. We present the NPRG method and
the approximation scheme (the DE) that we implement in
Sect.~\ref{nprgandde}. We then review the analysis of the Ising case
in Sect.~\ref{sec_Ising_review} presented in
Ref.~\cite{Balog:2019rrg}. Sect.~\ref{errorbars} is devoted to the
description of the methodology proposed to estimate error bars. In
Sect.~\ref{sec_results}, we give our determinations of the critical
exponents for various values of $N$, including the
physical cases $N=2, 3$ as well as the non-unitary cases $N=0$ and
$N=-2$. We also test large values of $N$ to compare with the large-N
results. Some technical details are addressed in Appendices.

\section{Non-Perturbative Renormalization Group and Derivative Expansion}
\label{nprgandde}
\subsection{The Non-Perturbative Renormalization Group}

We start with a brief review of the NPRG. It is based on Wilson's ideas of 
integrating first the highly
oscillating modes (i.e. those with a wave-vector larger than some
scale $k$) while the long-distance modes are frozen.

A convenient implementation of this consists in adding to the Euclidean action (or 
Hamiltonian) a
regulating term, quadratic in the fields and dependent on a momentum scale $k$ \cite{Polchinski:1983gv},
$S[\varphi]\to S[\varphi]+\Delta S_k[\varphi]$ with:
\begin{equation}
\label{deltaS}
 \Delta S_k[\varphi]=\frac 1 2 \int_{x,y}\varphi_a(x) R_k(x,y) \varphi_a(y),
\end{equation}
where $\int_x=\int d^dx$.  The regulating function $R_k$ is chosen to
be invariant under rotations and translations and therefore depends
only on $|x-y|$. Here and below, Einstein convention is adopted both
on sums over internal and space indices.  To properly regularize the
theory in the infrared, the Fourier transform $R_k(q)$ of $R_k(x-y)$
should:
\begin{itemize}
 \item be a smooth function of the modulus of the momentum $q$;
 \item behave as a ``mass square'' of order $k^2$ for long-distance modes:
   $R_k(q)\sim Z_k k^2$ for $q\ll k$, where $Z_k$ is a field renormalization 
factor to be specified below;
 \item go to zero rapidly when $q\gg k$ (typically faster than any power law).
\end{itemize}
With these properties the term (\ref{deltaS}) regularizes the theory
in the infrared without modifying the ultraviolet regime. One can then
define a scale-dependent partition function in the
presence of an arbitrary external source $J$
\cite{Wetterich:1992yh,Ellwanger:1993kk,Morris:1993qb}:
\begin{equation}
\label{regulatedgeneratingfunc}
{\cal Z}_k[J]= e^{W_k[J]}=\int \mathcal{D}\varphi\  e^{-S[\varphi]-\Delta 
S_k[\varphi]+\int_x J_a(x) \varphi_a(x)},
\end{equation}
where $W_k[J]$ is the Helmholtz free-energy or generating functional
of connected correlation functions. The Gibbs free-energy, or
scale-dependent effective action, is defined as the modified Legendre
transform of $W_k[J]$:
\begin{equation}
 \Gamma_k[\phi]=\int_x \phi_a(x) J_a(x) -W_k[J]-\Delta S_k[\phi].
\end{equation}
In the previous equation, $J$ is an implicit function of $\phi$,
obtained by inverting
\begin{equation}
  \phi_a(x)=\frac{\delta W_k}{\delta J_a(x)}.
\end{equation}
The theory is defined at a microscopic scale $\Lambda$
as the inverse of lattice spacing.
$\Gamma_k[\phi]$ is the generating functional of IR-regularized proper vertices defined
as
\begin{equation}
 \Gamma^{(n)}_{a_1\dots a_n}[x_1,\dots,x_n;\phi]
 =\frac{\delta^n \Gamma_k[\phi]}{\delta \phi_{a_1}(x_1)\dots \delta\phi_{a_n}(x_n)}.
\end{equation}
Here and below, we have omitted to indicate the $k$-dependence of the
regularized proper vertices to alleviate notation. As is well-known
only 1PI perturbative diagrams contribute to proper vertices. In
actual calculations, we will be interested in proper vertices
evaluated in a uniform field. We therefore define
\begin{equation}
 \Gamma^{(n)}_{a_1\dots a_n}(x_1,\dots,x_n;\phi)
 =\Gamma^{(n)}_{a_1\dots a_n}[x_1,\dots,x_n;\phi(x)]\big|_{\phi(x)\equiv\phi}.
\end{equation}
The Fourier transform of the vertices are defined as:
\begin{align}
\label{eqgamma_n}
&\Gamma^{(n)}_{a_1\dots a_n}(p_1,\dots,p_{n-1};\phi)\nonumber\\
&=\int_{x_1\dots x_{n-1}} \,\mathrm{e}^{i \sum_{m=1}^{n-1}x_m\cdot 
p_m}\Gamma^{(n)}_{a_1\dots a_n}(x_1,\dots,x_{n-1},0;\phi).
\end{align}
which only depend on $n-1$ independent wave-vectors because of
the invariance under translations.

The evolution of $\Gamma_k[\phi]$ with the RG time $t=\log(k/\Lambda)$  
\cite{Wetterich:1992yh,Ellwanger:1993kk,Morris:1993qb} can be easily obtained:
\begin{equation}
\partial_{t}\Gamma_{k}[\phi]=\frac{1}{2}\int_{x,y}\partial_{t}R_{k}(x-y)G_{aa}[x,y;\phi]
\label{wettericheq}
\end{equation}
where $G_{ab}[x,y;\phi]$ is the propagator in an arbitrary external
field $\phi(x)$, which has a matrix structure because of the internal
indices.  Here again, we omit to indicate the $k$-dependence of the
propagator to alleviate notations. The propagator can be obtained from
the two-point vertex in a standard way:
\begin{equation}
\int_{y} G_{ac}[x,y;\phi]\Big[\Gamma^{(2)}_{cb}[y,z;\phi]+ R_k(y-z)\delta_{cb}\Big]=\delta(x-z)\delta_{ab}
\end{equation}
The exact flow equation (\ref{wettericheq}) is a nonlinear functional equation.
From this functional equation, one can derive equations for
the various proper vertices.  For instance, evaluating (\ref{wettericheq})
in a uniform external field one deduces the exact equation for the effective potential
(or 0-point vertex in a uniform field $\phi$):
\begin{equation}
\label{eqpot}
 \partial_t U_k(\phi)=\frac 1 2 \int_q \partial_t R_k(q) G_{aa}(q;\phi).
\end{equation}
As for the potential, the
equation for the 2-point function in a uniform external field can be deduced by taking two functional
derivatives of (\ref{wettericheq}) and then evaluating in a uniform field. This gives, after Fourier transform:
\begin{align}
 \partial_t &\Gamma_{ab}^{(2)}(p;\phi)=\int_q \partial_t R_k(q) G_{mn}(q;\phi)
 \Big\{-\frac 1 2 \Gamma_{abns}^{(4)}(p,-p,q;\phi)\nonumber\\
&+ \Gamma_{anl}^{(3)}(p,q;\phi)G_{lr}(p+q;\phi)\Gamma_{bsr}^{(3)}(-p,-q;\phi)
 \Big\}G_{sm}(q_\phi)
 \label{eqvertex2}
\end{align}
Similarly, one can deduce the equation for any vertex function. As is
well-known, it leads to an infinite hierarchy of coupled equations,
where the equation for $\Gamma^{(n)}$ depends on all the vertices up
to $\Gamma^{(n+2)}$.  As a consequence Eq.~(\ref{wettericheq}), or
equivalently, the infinite hierarchy for vertex functions cannot be
solved without approximations, in the most interesting cases.

The asset of Eq.~(\ref{wettericheq}) compared to other functional
equations is that it is well-suited to formulate approximations going
beyond perturbation theory. In particular,
\begin{itemize}
 \item It has a one-loop structure written exclusively in terms of running and 
regularized vertices extracted from $\Gamma_k$.
 \item It has a 1PI structure (only dressed 
1PI diagrams contribute).
 \item In Fourier space, only internal momenta $q\lesssim k$ contribute 
significantly to the flows of any vertex.
\end{itemize}
This structure is clearly visible in the equations (\ref{eqpot}) and
(\ref{eqvertex2}) for the effective potential and 2-point vertex and
one can easily see that the same property holds for any
vertex-function.

In the next paragraph we present the most studied approximation going beyond 
perturbation theory within the NPRG: the DE. It fully
exploits the specific properties of the NPRG that we just mentioned.

\subsection{The Derivative Expansion}
\label{DEgeneral}
The DE procedure consists in taking an \textit{ansatz} for the
effective action $\Gamma_k[\phi]$ in which only terms with a finite
number of derivatives of the fields appear. Equivalently, in Fourier
space, it corresponds to expanding all proper vertices in power series
of the momenta and truncating to a finite order.  This approximation
is only well-suited for studying the long-distance properties of the
system since higher momentum dependence are neglected. In fact, it
proved to be a good approximation scheme for $\mathbb{Z}_2$ and $O(N)$
models with a very good level of precision (see for example,
\cite{Berges:2000ew,Canet:2003qd,Canet:2002gs,Balog:2019rrg}).  One of
the reasons for the success of the DE in $O(N)$ models is that its
predictions for many universal quantities, including critical
exponents, become exact not only for $4-d\ll 1$ but also for
$d-2\ll 1$ (for $N>2$) (see, for example,
\cite{Berges:2000ew,Delamotte:2007pf}) and for any $d$ in the
large-$N$ limit \cite{DAttanasio:1997yph}. The DE is, at least, an
educated interpolation between well-known limits.

Since the original works on NPRG \cite{Wetterich:1992yh,Morris:1993qb}
it was argued that the exact equations have a dressed one-loop
structure where all propagators are regularized in the infrared,
ensuring the smoothness of the vertices as a function of momenta and
allowing such an expansion. Moreover, the integral in
Eq.~(\ref{wettericheq}) -- or its derivatives with respect to the
fields such as (\ref{eqpot}) or (\ref{eqvertex2})-- includes
$\partial_t R_k(q)$ in the numerator, which tends rapidly to zero when
$q\gtrsim k$. This implies that the integral over $q$ is dominated by the
range $q \lesssim k$.  A further progress was made in
Refs.~\cite{Blaizot:2005xy,Benitez:2011xx} where the regime of
validity of this approximation has been discussed. It was observed
that an expansion in all momenta (internal and external) gives
equations that couple only weakly to the regime of momenta $p\gg
k$. Accordingly, it makes sense for these equations to formulate the DE
approximation scheme that only applies to the calculation of vertices
and its derivatives for momenta that are smaller than $k$ or the
smallest mass in the problem. In the case of critical phenomena when
$k\to 0$ the regime of validity of the DE reduces to those quantities
dominated by zero momenta (as thermodynamic properties or critical
exponents).

The radius of convergence of this expansion depends on the 
model considered and on the regulating function $R_k$.
However, in models described by Ginzburg-Landau
Hamiltonians whose analytical continuation to the Minkowskian space gives
unitary models, the radius has been shown
to be of the order  $q_{radius}^2/k^2\simeq 4$--9 \cite{Balog:2019rrg} once an appropriate 
regulator is chosen
with very specific {\it a priori} criteria. On top of the previous specifications, one needs
to fix the scale associated with the normalization of the fields in such a way
that all correlation functions at momenta $q^2/k^2\lesssim 4$--9 behave as in the 
massive theory. In that case the convergence of the DE takes 
place as in the massive theory
and the dependence on the regulator becomes locally smallest.
In practice this requirement is implemented by using a ``Principle of 
Minimal Sensitivity'' (PMS)
\cite{Stevenson:1981vj,Canet:2002gs} that is explained in detail below.

Given that the integral in (\ref{wettericheq}) [or similarly the
integrals for vertex functions such as (\ref{eqpot}) or
(\ref{eqvertex2})] are dominated by internal momenta of order
$q\lesssim k$, each successive order in the DE has an error, in low
momenta properties of the theory, that is suppressed by a factor
$1/9$--$1/4$.  The radius of order $4$ to $9$ corresponds to the ratio
between the square of the smallest mass in the Minkowskian version of
the model, and the minimum energy of the 2-particle (or 3-particle)
state.  If the regulator is chosen properly, the error in the
calculation of correlation functions is reduced by a factor
$1/9$--$1/4$ at successive order of the DE.

The quality of most DE results is further improved
at low orders because of an independent reason, as explained in Ref.~\cite{Balog:2019rrg}.
Consider the 2-point function near the fixed point and define:
\begin{equation}
\gamma_{ab}(p;\phi)= \frac{\Gamma_{ab}^{(2)}(p;\phi)-\Gamma_{ab}^{(2)}(p=0;\phi)}{p^2}.
\end{equation}
When $p\gg k$ and/or $\phi\gg k^{\frac{d-2+\eta}{2}}$, there is a
physical scale that regulates the theory and the regulator can be
neglected. As a consequence, in this regime, the function has a
scaling behavior and behaves as
\begin{equation}
\gamma_{ab}(p;\phi)\sim p^{-\eta}\hat{\gamma}_{ab}\Big(\frac{\phi}{p^{(d-2+\eta)/2}}\Big).
\end{equation}
This means that, in the scaling regime, both the dependence on $p$ and
on $\phi$ are controlled by an exponent of the order of $\eta$. By
continuity, in the opposite regime $p\lesssim k$, the function
$\gamma_{ab}(p;\phi)$ must show a dependence on $p$ and $\phi$ of the
order of magnitude of $\eta$. This means that, for the 2-point
function, all corrections to the LPA (where all terms with derivatives
are depreciated, except for an unrenormalized term $(\partial\phi)^2$)
are suppressed by a factor of $\eta$ which, in many models, is very
small. As a consequence, all quantities that can be extracted from the
2-point function in a uniform magnetic field (such as the exponents
$\eta$ and $\nu$) are already very well estimated at order
$\mathcal O(\partial^2)$. This makes the convergence very fast in all
cases where the exponent $\eta$ is small.  It is important to stress
that this {\it does not} mean that the expansion parameter of the DE
is of order $\eta$.  This factor suppresses all corrections to LPA but
does not suppress successive orders of the DE which are only suppressed
by a factor $1/9$--$1/4$.

This analysis is applicable, in particular, in the important
$\mathbb{Z}_2$ and $O(N)$ models with $N\geq 1$.  This is consistent
with the fact that the DE shows a rapid apparent convergence at low
orders for $O(N)$ models.  In fact, the DE has been pushed with
success to orders $\mathcal{O}(\partial^4)$ \cite{Canet:2003qd} and
$\mathcal{O}(\partial^6)$ \cite{Balog:2019rrg} for the Ising
universality class, giving excellent results that improve
significantly with the order of the DE.  Below, it will be shown that
the quality of the results extends to all $O(N)$ models at order
$\mathcal{O}(\partial^4)$. A mention must be made to the appearance of
Goldstone modes in the broken phase of $O(N)$ models for $N\neq
1$. Naively, one could think that the analysis of
Ref.~\cite{Balog:2019rrg} does not apply (at least in the low
temperature phase) because of the existence of these zero-mass
modes. However, the expansion must be done by including the full
propagator that includes the regulating function. This gives a square
mass of the order of $R_k(0)$ to all modes, including the Goldstone
modes. Hence, in the same way as in the regulated theory, the critical
regime behaves as a massive theory in the $N=1$ case and additionally
both the critical regime and the low temperature phase behave as
massive theories for $k>0$ (even if the theory presents massless modes
when $k\to 0$).

We indicate here that there exist several exact RG equations, which
have different convergence properties in the DE.  For
example, the Wilson-Polchinski equation
\cite{Wilson:1973jj,Polchinski:1983gv} involves all 1-Particle
Reducible diagrams which generate the connected correlation functions,
may they be 1-Particle Irreducible (1PI) or not. This is at odds with
NPRG equations where only 1PI diagrams contribute.  As a consequence,
the radius of convergence is of order $q^2/k^2 \sim 1$. This implies
that the DE for the Wilson-Polchinksi flow has a control parameter of
order one, which explains why the DE gives much better results in the
NPRG formulation \cite{Berges:2000ew}, even at order
$\mathcal{O}(\partial^2)$, than in the Wilson-Polchinski's one
\cite{Bervillier:2005za} (as had also been observed in perturbation
theory \cite{Morris:1999ba}).

In the present work we will analyze only critical exponents (that are
universal).  We can therefore use as microscopic action a simple
Ginzburg-Landau model with Hamiltonian or Euclidean Action,
\begin{equation}
S[\phi]=\int_x \Big\{\frac{1}{2}\big(\partial_\mu\phi^a\big)^2+\frac r 2 
\phi^a\phi^a+
\frac{u}{4!}(\phi^a\phi^a)^2\Big\}.
\end{equation}
In order to implement the DE, one considers, at each order of the approximation,
the most general terms compatible with the
symmetries of a given universality class with a limited number of derivatives.  In 
the case of the critical regime of $O(N)$ models, we require invariance under space isometries and
under the internal $O(N)$ symmetry. To be explicit, in the $O(N)$
model, the lowest orders approximations are: \\
\indent $\bullet$ The
Local Potential Approximation (LPA) or order $\mathcal{O}(\partial^0)$
which consist in taking no derivative of the field except a bare,
unrenormalized, kinetic term:
\begin{equation}
\Gamma_k^{\partial^0}[\phi]=\int_x \Big\{ 
U_k(\rho)+\frac{1}{2}\big(\partial_\mu\phi^a\big)^2\Big\}.
\end{equation}
Here, the running effective potential $U_k(\rho)$ is an arbitrary
function of $\rho=\phi_a\phi_a/2$.

\indent $\bullet$ The $\mathcal{O}(\partial^2)$, which is the
next-to-leading order, consists in taking all the possible terms
compatible with the symmetries of the model and with at most
two derivatives. In this case the \textit{ansatz} reads:
\begin{equation}
	\Gamma_k^{\partial^2}[\phi]=\int_x \Big\{ 
U_k(\rho)+\frac{1}{2}Z_k(\rho)\big(\partial_\mu\phi^a\big)^2
	+\frac{1}{4}Y_k(\rho)\big(\partial_\mu\rho\big)^2 \Big\}.
\label{ansatz-order2}
\end{equation}
For $N=1$ the terms in $Z_k(\rho)$ and $Y_k(\rho)$ are equivalent and,
accordingly, only the $Z_k(\rho)$ function is included.

\indent $\bullet$ Along the same lines, the order $\mathcal{O}(\partial^4)$, which is the 
next-to-next-to-leading order,
gives rise to the \textit{ansatz}:
\begin{align}
\nonumber\Gamma_k^{\partial^4} &[\phi]=\int_x \Big\{  
U_k(\rho)+\frac{1}{2}Z_k(\rho)\big(\partial_\mu\phi^a\big)^2
	+\frac{1}{4}Y_k(\rho)\big(\partial_\mu\rho\big)^2
\\	\nonumber
&+\frac{W_1(\rho)}{2}\big(\partial_\mu\partial_\nu \phi^a\big)^2  \nonumber
+\frac{W_2(\rho)}{2}\big(\phi^a \partial_\mu\partial_\nu \phi^a\big)^2 \\
& \nonumber + W_3(\rho)\partial_\mu\rho\partial_\nu\phi^a\partial_\mu\partial_\nu \phi^a
+\frac{W_4(\rho)}{2}
\phi^b\partial_\mu\phi^a\partial_\nu\phi^a\partial_\mu\partial_\nu \phi^b \\
\nonumber &
+ \frac{W_5(\rho)}{2}
\varphi^a\partial_\mu\rho\partial_\nu\rho\partial_\mu\partial_\nu 
\varphi^a
+\frac{W_6(\rho)}{4}\Big(\big(\partial_\mu\varphi^a\big)^2\Big)^2 \\
\nonumber & +\frac{W_7(\rho)}{4}\big(\partial_\mu\phi^a\partial_\nu\phi^a\big)^2
+\frac{W_8(\rho)}{2}
\partial_\mu\phi^a\partial_\nu\varphi^a\partial_\mu\rho\partial_\nu\rho
\\ &
+\frac{W_9(\rho)}{2}\big(\partial_\mu\varphi^a\big)^2 \big(\partial_\nu\rho\big)^2 
+\frac{W_{10}(\rho)}{4} \Big(\big(\partial_\mu \rho)^2\Big)^2  \Big\}.
\label{ansatz-order4}
\end{align}
As for the order $\mathcal{O}(\partial^2)$, there are many terms in 
the $O(N)$ case
at order $\mathcal{O}(\partial^4)$
that are identical in the $\mathbb{Z}_2$ case. Indeed, in the $\mathbb{Z}_2$ 
case there are only
three independent terms \cite{Canet:2003qd} (see below) with four derivatives.

\indent $\bullet$ The order $\mathcal{O}(\partial^6)$ has only been analyzed in 
the
$\mathbb{Z}_2$ universality class \cite{Balog:2019rrg}.  In that case, the 
ansatz for $\Gamma_k[\phi]$ reads: 
\begin{eqnarray}
& & \Gamma_k^{\partial^6,\mathbb{Z}_2}[\phi]= \int_x \Big\{U_k(\phi) + \tfrac{1}{2} 
Z_k(\phi) (\partial_\mu\phi)^2 \nonumber\\
&& + \tfrac{1}{2} W^a_k(\phi)(\partial_\mu\partial_\nu\phi)^2 + 
\tfrac{1}{2} \phi W^b_k (\phi)(\partial^2\phi)(\partial_\mu\phi)^2 \nonumber\\
&& + \tfrac{1}{2}  W^c_k 
(\phi)\left((\partial_\mu\phi)^2\right))^2+ \tfrac{1}{2} \tilde X^a_k 
(\phi)(\partial_\mu\partial_\nu\partial_\rho\phi)^2 \nonumber\\
&& + \tfrac{1}{2} \phi \tilde X^b_k 
(\phi)(\partial_\mu\partial_\nu\phi)(\partial_\nu\partial_\rho\phi)(\partial_\mu
\partial_\rho\phi) \nonumber\\
&& + \tfrac{1}{2}  \phi\tilde 
X^c_k(\phi)\left(\partial^2\phi\right)^3
+\tfrac{1}{2}  \tilde X^d_k (\phi)\left(\partial^2\phi\right))^2 
(\partial_\mu\phi)^2\nonumber \\
&& + \tfrac{1}{2} \tilde X^e_k (\phi)(\partial_\nu\phi)^2 
(\partial_\mu\phi)(\partial^2\partial_\mu\phi)
+ \tfrac{1}{2} \tilde X^f_k (\phi)(\partial_\rho\phi)^2 
(\partial_\mu\partial_\nu\phi)^2 \nonumber\\
&& + \tfrac{1}{2}  \phi\tilde X^g_k 
(\phi)\left(\partial^2\phi\right)\left((\partial_\mu\phi)^2\right)^2
+ \tfrac{1}{96} \tilde{X}^h_k (\phi) \left((\partial_\mu\phi)^2\right)^3 \Big\}.
\label{ansatz-order6-complet}
\end{eqnarray}
For $O(N)$ models at order $\mathcal{O}(\partial^6)$, instead of eight
independent functions $X^i(\rho)$ corresponding to terms with six derivatives
as in the $N=1$ case, one must introduce 48 independent functions
of $\rho$ whose treatment would be a formidable task.

At a given order of the DE, the flow of the various functions is
obtained by inserting the corresponding {\it ansatz} in
Eq.~(\ref{wettericheq}) and expanding and truncating the
right-hand-side on the same functional subspace. For instance, in
order to deduce the equation for the effective potential at order
$\mathcal{O}(\partial^2)$ one must insert in the exact equation
(\ref{eqpot}) the propagator obtained from the 2-point vertex extracted
from the {\it ansatz} (\ref{ansatz-order2}):
\begin{align}
\label{eqgamma2DE2}
\Gamma_{ab}^{(2)}(p;\phi)
&=\delta_{ab}\big(U_k'(\rho)+Z(\rho)p^2\big)\nonumber\\
&+\phi_a \phi_b \big(U_k''(\rho)+\frac{1}{2}Y(\rho)p^2\big)+\mathcal{O}(p^4),
 \end{align}
 As for the potential, the equation for $Z_k(\rho)$ or $Y_k(\rho)$ can
 be obtained from the equation for the 2-point function in a uniform
 external field.  In order to do so, one must express those functions
 in terms of the vertices (or its derivatives) in a uniform field. For
 example,
\begin{equation}
Z_k(\rho)=\frac{1}{N-1}\Big(\delta_{ab}-\frac{\phi_a\phi_b}{2\rho}\Big)\partial_
{p^2}\Gamma_{ab}^{(2)}(p;\phi)\vert_{{\bf p}=0}.
\end{equation}
It is obvious from the previous expression that the $N=1$ case must be treated 
separately. This is a manifestation of the fact that for $N=1$ the terms in the
effective action including $Z(\rho)$ and $Y(\rho)$ are 
different representations of an identical term. A similar procedure can be
implemented at any order of the DE.

The flow equations for the various functions have been obtained in the
past at order $\mathcal{O}(\partial^2)$ (see, for example,
\cite{VonGersdorff:2000kp}) and, for $N=1$, at order
$\mathcal{O}(\partial^4)$ \cite{Canet:2003qd,Balog:2019rrg}. We
obtained our equations for arbitrary $N$ at order
$\mathcal{O}(\partial^4)$ by implementing a Mathematica code.  We
verified that our equations reduce to previously known
$\mathcal{O}(\partial^2)$ equations when $\mathcal{O}(\partial^4)$
terms are neglected. We also verified that we recovered previous
$\mathcal{O}(\partial^4)$ results for $N=1$ in the corresponding
limit.  We point out that in this work, like in the previous
$\mathcal{O}(\partial^6)$ work of \cite{Balog:2019rrg}, we implement
in the flow equations a strict polynomial expansion in momenta at the
considered order of the DE. For instance, one of the contributions to
the flow of $\Gamma^{(2)}(p)$ at order $\mathcal O(\partial^4)$ involves the
product of two $\Gamma^{(3)}$ functions, see Eq.~(\ref{eqvertex2}). At
this order of the DE, these two functions are polynomials of order 4
in their momenta, and their product therefore involves up to 8 powers
of the momenta. In our implementation of the DE at $O(\partial^4)$, we
drop such terms, as well as other terms which contain more than 4
powers of the momenta. This differs from more standard implementations
of the DE \cite{VonGersdorff:2000kp,Canet:2003qd} where all terms are
kept in the flows even though many other terms of order 6 and 8 have
been neglected \footnote{We verified explicitly that both versions of
  the DE give results that are compatible within error bars at
  order $O(\partial^2)$.}. Of course, we use the same procedure for
all the flows that we consider.  We verified that our Mathematica code
recovers properly both versions of the equations.  In practice, our
implementation of the DE yields much simpler expressions for the flows
than those obtained with the standard implementation of the DE. They
are moreover probably much better under control numerically at order 4
where the flows of some functions involve the product of four
$\Gamma^{(3)}$ functions.  The details of the numerical solution of
the equations is presented in Appendix~\ref{Ap:NumParam}.

At criticality ---the regime on which we focus in this article--- the
system is scale invariant.  To reach the critical regime typically
requires to fine-tune one bare coupling in the initial condition for
the flow equations.  The Ward identities for scale invariance in the
presence of the infrared regulator $\Delta S_k$ are equivalent to a
{\it fixed point} condition on the flow of $\Gamma_k$, that is,
$\partial_t\Gamma_k=0$ when it is expressed in terms of dimensionless
and renormalized quantities \cite{delamotte2016scale}.  More
precisely, one defines renormalized and dimensionless fields and
coordinates by
\begin{align}
\tilde x&=k x,\\
\phi^a(x)&=k^{(d-2)/2}Z_k^{-1/2}\tilde \phi^a(\tilde x), \\
\rho(x)&=k^{(d-2)}Z_k^{-1}\tilde \rho(\tilde x).
\end{align}
and functions $\tilde{F}(\tilde\rho(\tilde{x}))$:
\begin{equation}
F(\rho)=k^{d_F}Z_k^{n/2}\tilde{F}(\tilde\rho)
\end{equation}
where $F(\rho)$ is any function appearing in the {\it ansatz} for
$\Gamma_k$: $U_k(\rho), Z_k(\rho), \cdots, W^{10}(\rho)$, $d_F$ is the
canonical dimension of $F$ and $n$ the number of fields $\phi^a$ that
multiply $F$ in $\Gamma_k$. The field renormalization factor $Z_k$
which appears in previous equations is related to the function
$Z_k(\rho)$ in the following way. We first define the renormalized
equivalent of $Z_k(\rho)$ by the relation
$Z_k(\rho)=Z_k\tilde{Z}_k(\tilde\rho)$. The renormalization factor
$Z_k$ is then defined by the (re)normalization condition:
$\tilde Z_k(\tilde\rho_0)=1$ for a fixed value of $\tilde\rho_0$.  The
running anomalous dimension $\eta_k$ is then defined by $\eta_k= -
\partial_t\log Z_k$. At the fixed point,
it becomes the actual anomalous dimension $\eta$ 
\cite{Berges:2000ew}.

\section{Review of previous Derivative Expansion results for the $N=1$ case}
\label{sec_Ising_review}
We now consider results in the Ising universality class (corresponding
to the $N=1$ case). This universality class has been studied many
times at LPA and order $\mathcal{O}(\partial^2)$
\cite{Wegner:1972ih,Hasenfratz86,Tetradis94,Wetterich:1992yh,Morris:1994ie,Morris:1997xj,
  Seide99,Litim02,Canet:2002gs,Berges:2000ew,VonGersdorff:2000kp,Delamotte:2007pf}
and even at order $\mathcal{O}(\partial^4)$ \cite{Canet:2003qd}.
However, depending on the authors, slightly different flow equations
have been considered and in some cases, on top of the DE, the 
functions $U_k(\phi)$, $Z_k(\phi)$ and $W_k^i(\phi)$ have been
replaced by their Taylor expansion in $\rho$ truncated at finite
order.  In a recent article \cite{Balog:2019rrg}, the order
$\mathcal{O}(\partial^6)$ has been analyzed and the critical exponents
$\eta$ and $\nu$ were compared to the very precise results coming
from the CB \cite{ElShowk:2012ht,El-Showk:2014dwa,Kos:2014bka}.  This
calculation has allowed for a quantitative analysis of the error of
the DE up to order 6 and has suggested a methodology to estimate the
error bars. We show in detail in Sect.~\ref{errorbars} how to
implement this analysis in the $O(N)$ case. The information obtained
for $N=1$ will be an important guide when estimating error bars in the
more general $O(N)$ case.  In the present section we review the $N=1$
results. In Sect.~\ref{omegaN1} we extend them at order
$\mathcal{O}(\partial^4)$ for the exponent $\omega$ (related to
corrections to scaling).

At this point, it is important to stress that the DE ---like any
approximation scheme--- introduces a spurious dependence of the
critical exponents on the regulating function $R_k$ mentioned in
Sect.~\ref{DEgeneral}. In the DE the role of the regulator is more
important than in other approximations because the mere formulation of
the approximation {\it requires} that we have introduced the
regulator. That is, the approximation is only justified for momenta
smaller than two to three times $k$, the scale of the regulator.  This
is at odds with other approximation schemes which do not treat only a
limited range of momenta (as, for
example,~\cite{Blaizot:2005xy,Benitez09,Benitez:2011xx}).

For each family of regulators we analyzed the dependence of critical
exponents on the regulating function on the overall scale
$\alpha$. For the considered regulators it turns out to be the most
important dependence on the regulating function. In order to fix this
scale, we use the ``Principle of Minimal Sensitivity" (PMS)
\cite{Stevenson:1981vj,Canet:2002gs}. The underlying rationale is
that, in the exact theory, the exponents do not depend on the overall
scale of the regulator, and therefore an optimal choice of $\alpha$ is
obtained when the physical results depend least on this parameter. In
many cases, the exponents as a function of $\alpha$ present a local
extremum and in those cases the $\alpha_{PMS}$ is just the value
corresponding to this extremum.

In a second step, one can change the shape of the regulating function
and see how the PMS results are spread.  In Ref.~\cite{Balog:2019rrg}
three families of regulators were employed:
\begin{subequations}
\begin{align}
W_k(q^2) &= \alpha Z_k k^2 \, y/(\exp(y) - 1) \label{regulator-wetterich}\\ 
\Theta^n_k(q^2) &= \alpha Z_k k^2 \, (1-y)^n \theta(1-y)   \;\;\;  
\label{regulator-theta}\\
E_{k}(q^2) &= \alpha Z_k k^2 \,\exp(-y)  \label{regulator-exp}
\end{align}
\label{regulators}
\end{subequations}
where $y=q^2/k^2$.

The regulator (\ref{regulator-wetterich}) with $\alpha=1$ was proposed
by Wetterich \cite{Wetterich:1992yh} and the $\alpha$-dependence of
physical quantities such as exponents was studied in
\cite{Canet:2002gs}, \cite{Canet:2003qd} and \cite{Balog:2019rrg}.  It
turns out that fixing the prefactor by the PMS procedure improves
significantly the results of the DE compared to the standard value
$\alpha=1$.  Being smooth, the DE can be applied with this regulating
function at any order.  Another convenient regulator was proposed by
Litim \cite{Litim02}.  It is non-analytic but smoother than the sharp
cut-off \cite{Wegner:1972ih} commonly employed with the
Wilson-Polchinski equations \cite{Wilson:1973jj,Polchinski:1983gv}
(see below).  It corresponds to the $\Theta^n_k$ regulator defined in
Eq.~(\ref{regulator-theta}) with $n=1$ and $\alpha=1$.  It allows for
the analytic calculation of many integrals involved in the LPA flows
and there are strong indications that this is the optimal choice at
the LPA order \cite{Litim02,Morris:2005ck}.  However, since it is
non-analytic, it is not well-suited for a systematic expansion in
momenta, as is done in the DE.  Moreover, it turns out that it is not
optimal at order $\mathcal{O}(\partial^2)$ \cite{Canet:2002gs} and is
incompatible with the DE at order $\mathcal{O}(\partial^4)$ due to the
non- analyticities it induces in the flows.  At any finite order of
the DE, regulators of the family (\ref{regulator-theta}) can be used
under the condition that their index $n$ is large enough to keep the
flows smooth-enough for the various considered functions to be
well-defined.  Another smooth regulator used in
\cite{Tissier_2012,Balog:2019rrg} corresponding to
expression~(\ref{regulator-exp}) will be considered below.  In that
case, also, it has been observed that the PMS turns out to be an
efficient optimization procedure \cite{Balog:2019rrg} for $N=1$ and,
as shown below, this is also true for more general $O(N)$ models.

Each regulating function studied in Ref.~\cite{Balog:2019rrg}
yielded very similar results once the overall scale $\alpha$ is fixed by the PMS.
It must be mentioned, however, that in the literature, other
regulators have been also considered giving results of lower quality. In 
particular, the sharp cut-off was employed by Wegner and Houghton at order
LPA \cite{Wegner:1972ih} long time before the modern implementation of NPRG. The sharp
cut-off corresponds to the regulating function
\begin{equation}
S_k(q^2)=\left\lbrace
\begin{array}{ll}
 \infty \hspace{.3cm} &\mathrm{if}\, q<k\\
 0 &\mathrm{if}\, q>k
\end{array}
\right.
\end{equation}
The strong non-analyticities induced by this regulator in the flows do
not allow for the implementation of the DE beyond the LPA
\footnote{Let us note, however, that Morris implemented a similar
  momentum scale expansion \cite{Morris:1995af} for this regulator.}.
Power law regulators have also been studied by Morris
\cite{Morris:1994ie,Morris:1997xj}. They however yield relatively poor
results at LPA and $\mathcal{O}(\partial^2)$ probably because of two
independent reasons. First, the large momentum region, which is beyond
the radius of convergence of the DE, is only suppressed in the
integrals involved in the flows as power laws at odds with the
regulators (\ref{regulator-wetterich}), (\ref{regulator-theta}) and
(\ref{regulator-exp}).  Second, being non-analytic at $q=0$, the
convergence properties of the DE are not controlled by the small
parameter discussed before.

\begin{figure}
     \begin{picture}(216,165)
      \put(0,0){\includegraphics[width=216pt]{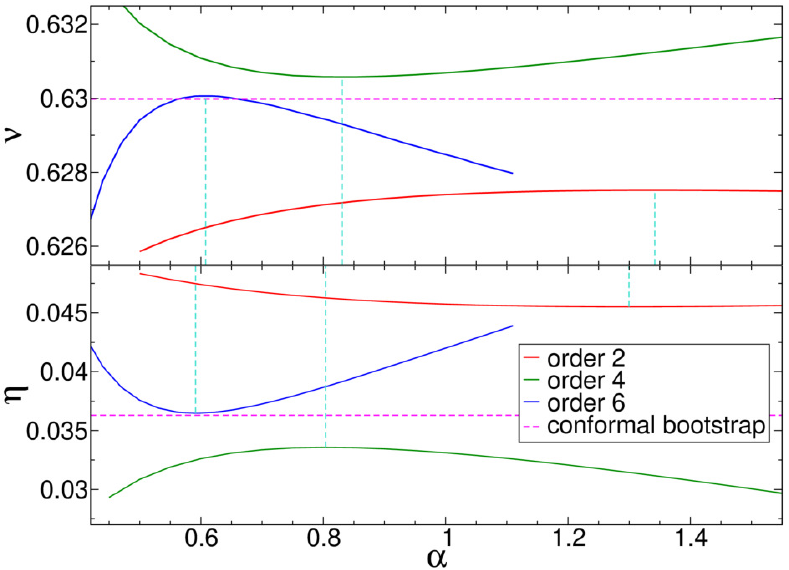}}
     \end{picture}
     \label{fig_opt}
     \caption{ Dependence of the critical exponents $\nu(\alpha)$ and
       $\eta(\alpha)$ with the coefficient $\alpha$ for different
       orders of the DE (figure from Ref.~\cite{Balog:2019rrg}). LPA
       results do not appear within the narrow ranges of values chosen
       here (see Table~\ref{table_suppl}).  }
     \label{pms}
\end{figure}

Let us discuss as an example the results obtained in \cite{Balog:2019rrg} with 
the regulator (\ref{regulator-exp}).
In Fig.~\ref{pms} (from Ref.~\cite{Balog:2019rrg}) the dependence of the critical exponents $\nu(\alpha)$
and $\eta(\alpha)$
with the coefficient $\alpha$ is represented for different orders of the DE.
At each order, the curves for the exponent exhibit a maximum or a 
minimum at some value $\alpha_{\rm PMS}$.
When increasing the order of the DE, the extrema
alternate between being a maximum and a minimum (this is true for both exponents). Following the PMS criteria, 
the values $\nu(\alpha_{\rm PMS})$ and $\eta(\alpha_{\rm PMS})$ can be selected 
as best estimates.
Moreover, given that the concavities of the curves alternate, these results are 
those that gives the fastest apparent convergence because
they reduce the difference between two consecutive orders. At a given order of
the DE, $\alpha_{\rm PMS}^{(\nu)}$ and $\alpha_{\rm PMS}^{(\eta)}$ are close but 
different, and their difference decreases with the
order of the DE,  see Fig.~\ref{pms}.

\begin{table}[]
\caption{\label{table_suppl} Raw results of the DE from various previous references for 
the Ising critical exponents $\nu$ and $\eta$ in $d=3$ obtained with various 
regulators. When a value of $\alpha$ different from PMS is employed, this is explicitly indicated.
The numbers in parentheses for DE results give the distance of the results to 
the CB \cite{Kos:2014bka} values given
here as the almost exact reference.}
\begin{ruledtabular}
\begin{tabular}{llll}
      & \hspace{-0.5cm}regulator          &  $\nu$         &  $\eta$ \\
\hline
LPA   &$W$ \cite{Canet:2002gs,Balog:2019rrg}  & 0.65059(2062)          &  0            \\
      &$\Theta^1$ \cite{Litim02}           & 0.64956(1956)          & 0               \\
      &$\Theta^3$ \cite{Balog:2019rrg}     & 0.65003(2006)          &  0                \\
      &$\Theta^4$ \cite{Balog:2019rrg}     & 0.65020(2023)          &  0            \\
      &$\Theta^8$ \cite{Balog:2019rrg}     & 0.65056(2059)          &  0            \\
      &$E$ \cite{Balog:2019rrg}            & 0.65103(2106)          &  0      \\           
      &$E$ \cite{Balog:2019rrg}            & 0.65103(2106)          &  0      \\           
      &Power-law \cite{Morris:1997xj}      & 0.66                   &  0 \\
      &$S$ \cite{Hasenfratz86}      & 0.687                   &  0 \\
      \hline   
$O(\partial^2)$
     &$W$ \cite{Balog:2019rrg}             & 0.62779(218)          &  0.04500(870)  \\
     &$W$ ($\alpha=1$) \cite{Seide99}      & 0.6307                &  0.0467     \\
     &$\Theta^1$ \cite{Canet:2002gs}       & 0.6260                &  0.0470               \\
     &$\Theta^2$ \cite{Balog:2019rrg}      & 0.62814(183)          &  0.04428(798)  \\
     &$\Theta^3$ \cite{Balog:2019rrg}      & 0.62802(195)          &  0.04454 (824)  \\
     &$\Theta^4$ \cite{Balog:2019rrg}      & 0.62793(204)          &  0.04474(844)  \\
     &$\Theta^8$ \cite{Balog:2019rrg}      & 0.62775(222)          &  0.04509(879)  \\
     &$E$ \cite{Balog:2019rrg}             & 0.62752(245)          &  0.04551(921)  \\
     &Power-law \cite{Morris:1997xj}       & 0.618                 &  0.054 \\
     \hline
$O(\partial^4)$
      &$W$ \cite{Balog:2019rrg}          & 0.63027(30)           & 0.03454(176)  \\
      &$W$ (field expansion)\cite{Canet:2003qd}          & 0.632           & 0.033  \\
      &$\Theta^3$ \cite{Balog:2019rrg}   & 0.63014(17)           & 0.03507(123)    \\
      &$\Theta^4$ \cite{Balog:2019rrg}   & 0.63021(24)           & 0.03480(150)  \\
      &$\Theta^8$ \cite{Balog:2019rrg}   & 0.63036(39)           & 0.03426(204)   \\
      &$E$  \cite{Balog:2019rrg}         & 0.63057(60)           & 0.03357(272)  \\
\hline
$O(\partial^6)$
      &$W$  \cite{Balog:2019rrg}         & 0.63017(20)          & 0.03581(49)  \\
      &$\Theta^4$ \cite{Balog:2019rrg}   & 0.63013(16)          & 0.03591(39)  \\
      &$\Theta^8$ \cite{Balog:2019rrg}   & 0.63012(15)          & 0.03610(20)  \\
      &$E$  \cite{Balog:2019rrg}         & 0.63007(10)          & 0.03648(18)  \\
      \hline
CB \cite{Kos:2014bka}   &     & 0.629971(4) 	    & 0.0362978(20) 
\end{tabular}
\end{ruledtabular}
\end{table}

The values of the exponents at $\alpha_{\rm PMS}$ converge very fast
to values very close to the CB quasi-exact ones.  In
most cases they also alternate around these values. The only exception
is at order $\mathcal{O}(\partial^6)$.  At this order, the optimal
value of $\nu$ `crosses' the CB values, but at this
order PMS results coincide with CB values up to three
or four significant digits, see Fig.~\ref{pms} and
Table~\ref{table_suppl}.  It is possible that this property is exact
for correlation functions [as seen for $N=1$ up to order
$\mathcal{O}(\partial^6)$ \cite{Balog:2019rrg}] but is only an
approximation for critical exponents that are not directly related to
a single correlation function.  Surprisingly, the local curvature at
$\alpha_{\rm PMS}$ increases with the order of the DE. That could
indicate that for generic values of $\alpha$ the convergence of the DE
could be doubtful but, at PMS, the exponents seem to converge to the
best estimates in the literature.  The increase of curvature at
$\alpha_{\rm PMS}$ and the accompanying faster variations of exponent
values with $\alpha$ when increasing the order of the DE imply that it
is crucial to work with the optimal values given by the PMS criteria,
that is $\nu(\alpha_{\rm PMS}^{(\nu)})$ and
$\eta(\alpha_{\rm PMS}^{(\eta)})$.

Once the parameter $\alpha$ is fixed with the PMS procedure, the speed of 
convergence is in agreement with the considerations about the radius
of convergence of the DE at criticality. That is, the amplitude of the 
oscillations of the optimal values considered as functions of the order of the DE
decreases typically by a factor that is consistent with the convergence 
estimate at each successive order (4 to 9, see Table~\ref{table_suppl}).
Moreover, for each exponent, the dispersion of
values (over all regulators studied) 
typically also decreases by similar factors when going from one order to the 
next.
This can also be interpreted as a manifestation of the radius of convergence of 
the DE, see Table \ref{table_suppl}.

After a rather extensive exploration of different regulators, the 
authors of \cite{Balog:2019rrg} have conjectured the existence,
for a given exponent and a given order of the DE, of an absolute extremum value,
(an absolute maximum or an absolute minimum, depending on the exponent and the order 
considered) that cannot be passed by any regulator. As we discuss below, 
this general conjecture seems not to
be fulfilled for all $O(N)$ models and all exponents. However, in many important 
cases, at least up to order $\mathcal{O}(\partial^4)$
and for exponents $\nu$ and $\eta$, it seems to be correct.
We discuss this point in the next section and explain
how it can be used to improve the estimate of critical exponents.

\section{Expansion parameter and error bars}
\label{errorbars}
In the present section, we exploit the existence of a small expansion
parameter in the DE of $O(N)$ models to estimate the error bars for
various critical exponents. As explained in Ref.~\cite{Balog:2019rrg}
and reviewed in the previous sections, when calculating vertex
functions or their derivatives at zero momenta, the DE is controlled
by a small parameter of order $1/9$--$1/4$.  This leads to a
well-grounded estimate of error bars that can be employed in general
models. We discuss and implement them concretely below both for the
Ising universality class and for general $O(N)$ models.

\subsection{A first estimate of error bars}
\label{pessimistic}

Let us first discuss a generic estimate of
error bars within the DE (at least for models where there is a unitary Minkowskian 
extension). Consider a physical quantity $Q$ that we aim at computing.
The procedure is simple:
\begin{itemize}
 \item For a given regulator family and at a given order of the DE, we choose as value of 
$Q$ the one corresponding to the $\alpha$ determined by
 PMS procedure.
 \item When comparing among different families of regulating functions, without 
further information, it is reasonable to choose the
 value at the center of the range of values for $Q$ obtained for the considered 
regulators. Let us call $\bar Q^{(s)}$ this estimate
 at order $\mathcal{O}(\partial^{s})$.
 \item Having determined the estimates $\bar Q^{(s)}$ at various orders, we 
choose as first error estimate at
 order $\mathcal{O}(\partial^s)$, $\bar\Delta Q^{(s)}=|\bar Q^{(s)}-\bar 
Q^{(s-2)}|/4$.  The 1/4 corresponds to the more conservative estimation of the small
parameter. Indeed, dividing by four in many cases turns out to be
 a pessimistic choice. Nevertheless, without further information, it is 
convenient to choose pessimistic error bars.
\end{itemize}
Observe that this procedure does not lead to an estimate of error bars
at order LPA because it requires at least two consecutive orders.  It
is also interesting to observe that it can be employed for any
physical observable (which can be extracted from a vertex or its
derivatives at zero momenta). In Table~\ref{N1estimatesanderrors} the
results of the present analysis are presented for the exponents $\nu$
and $\eta$ in the Ising universality class given in
Ref.~\cite{Balog:2019rrg}.  Comparing with the results of the CB, one
observes that the DE estimates seem to converge to the quasi-exact
values and that estimated error bars are correct (or, more precisely,
somewhat pessimistic).

On top of these estimates of error bars, it is necessary to take into account 
the dependence of the results among the various families of regulators which is 
an independent source of errors (in most cases much smaller than the one that we 
just considered). Moreover, we should have in mind that these estimates are
typically pessimistic but can become too optimistic in the exceptional case where
two consecutive orders of the DE accidentally cross. This possibility can
be avoided by considering a more typical (and pessimistic) estimate in the
case of exceptional ``crossings''. We discuss these independent sources of error in 
Sect.~\ref{err_reg}. 

Before considering this point, it is important to stress that given
our knowledge of the actual behavior of exponents at the various
orders in the Ising universality class, one can test in this case the
quality of the proposed estimate of error bars. This information can
be used to improve the estimate of central values and error bars as
explained in the next subsection. This will be exploited in
Sect.~\ref{sec_results} in the analysis of other $O(N)$ models at
order $\mathcal{O}(\partial^{4})$.

\begin{table*}
\caption{Analysis of error bars at orders $\mathcal{O}(\partial^0)$ (LPA) to 
$\mathcal{O}(\partial^6)$.
Raw data extracted of \cite{Balog:2019rrg}. See text for the precise definitions 
of various possible central values
and error bars. CB \cite{Kos:2014bka} values are also given for 
comparison.}
\label{N1estimatesanderrors} 
\begin{ruledtabular}
\begin{tabular}{lllllllllllll}
D.E. & $\bar\nu$ & $\bar\Delta\nu$ & $\tilde\nu$ & $\tilde\Delta\nu$ & 
$\Delta_{reg}\nu$ & $\Delta\nu$ & $\bar\eta$ & $\bar\Delta\eta$ & $\tilde \eta$ & $\tilde 
\Delta\eta$ & $\Delta_{reg}\eta$ & $\Delta\eta$ \\
LPA            &0.65030  &  --      &   0.64956  & --      & 0.00147 & -- &  0       
& --           & 0       & --        &0    &--   \\
$O(\partial^2)$&0.62783  &  0.00562 &   0.63082  & 0.00268 & 0.00062 & 0.00268 &  0.04490 
& 0.01122 & 0.03875 & 0.00554   &0.00123 & 0.00554 \\
$O(\partial^4)$&0.63036  &  0.00063 &   0.62989  & 0.00025 & 0.00043 & 0.00025 & 0.03432 
& 0.00264 & 0.03622 & 0.00115   &0.00150 & 0.00115 \\
$O(\partial^6)$&0.63012  &  0.00006 &   --       & --      & 0.00010 & 0.00016 & 0.03615 
& 0.00046 & 0.03597 & 0.00018   & 0.00067 & 0.00113 \\
\hline
CB  & 0.629971(4) &  & &	&  &   &  0.0362978(20) & 
\end{tabular}
\end{ruledtabular}
\end{table*}

\subsection{Improving the estimate of central values and error bars}
\label{improverror}
As explained in the previous section, in most cases the concavity of
the curve of exponents as a function of $\alpha$ alternates. Moreover,
the results obtained at a given order of the DE do not intersect with
the previous one.  As a consequence, in those cases, choosing the PMS
also leads to the fastest apparent convergence by reducing the
difference of critical exponents estimates in consecutive orders. In
those cases we also have strong reasons to believe that, up to that
order, the DE gives alternating bounds (upper or lower) of critical
exponents. As a consequence, the estimate $\bar Q^{(s)}$ is clearly
not the optimal choice and the extremum among the values obtained via
PMS for various reasonable regulators seem to be a much more
reasonable estimate. Let us call this extremum $Q_{ext}^{(s)}$.

Note, however, that this estimate does not fully exploit the information 
that
DE expansion, in those cases, are bounds. One then expects the exponent to lie
in-between the results obtained in
two consecutive orders of the DE. For example, for the $N=1$ exponent $\nu$ at 
order $\mathcal{O}(\partial^{4})$ one would expect
that the actual value of the exponent lies in the interval 
$[\nu_{ext}^{(2)},\nu_{ext}^{(4)}]$.
The value $\nu_{ext}^{(4)}=0.63014$ is not an optimal estimate of the exponent 
at that order because
it is in the border of the interval of expected values. Moreover, given the fact 
that two consecutive orders give alternating errors
that are reduced by a factor $1/9$--$1/4$, one expects the actual value of the 
exponent to be
closer to $\nu_{ext}^{(4)}$ than to $\nu_{ext}^{(2)}$. Taking into account these 
considerations,
when the values $Q_{ext}^{(s)}$ for some quantity are expected to be extrema, an 
improved estimate corresponds to shifting $Q_{ext}^{(s)}$ towards the center of interval
between $Q_{ext}^{(s)}$ and $Q_{ext}^{(s-2)}$ by 
$|Q_{ext}^{(s)}-Q_{ext}^{(s-2)}|/8$ and consider as error
estimate $\tilde \Delta Q^{(s)}=|Q_{ext}^{(s)}-Q_{ext}^{(s-2)}|/8$ \footnote{For a
radius of convergence $a$, we would shift by $|Q_{ext}^{(s)}-Q_{ext}^{(s-2)}|/(2a)$. We
choose the most conservative radius of convergence $a=4$.}. This new 
estimate of central value
will be called $\tilde Q^{(s)}$ and its explicit expression is:
\begin{equation}
 \tilde Q^{(s)}=\frac 1 8 \Big(7\, Q_{ext}^{(s)}+Q_{ext}^{(s-2)}\Big)
\end{equation}

For example, in the same example as before, it is reasonable to give as estimate 
of $\nu$ at
order $\mathcal{O}(\partial^{4})$ the following improved estimate of central values
and error 
bars:
\begin{align}
 \tilde \nu^{(4)}&=\nu_{ext}^{(4)}-|\nu_{ext}^{(2)}-\nu_{ext}^{(4)}|/8=0.62989
 \nonumber\\
 \tilde\Delta \nu^{(4)}&=|\nu_{ext}^{(2)}-\nu_{ext}^{(4)}|/8=0.00025
\end{align}

These improved estimates are reasonable (see
Table~\ref{N1estimatesanderrors}) as long as we have strong reasons to
believe that the DE gives, at this order, a bound for the considered
physical quantity $Q$. A necessary condition for this is that two
consecutive orders of the DE give results for the various families of
regulators that are disjoints and that, for any regulator, the
considered quantity represented as a function of $\alpha$ shows the
appropriate convexity. Among the results obtained for $N=1$ there is a
single exception. The results of orders $\mathcal{O}(\partial^4)$ and
$\mathcal{O}(\partial^6)$ overlap for the exponent $\nu$: the $\nu$
exponent obtained with regulator $W$ at order
$\mathcal{O}(\partial^6)$ is larger than the one obtained with
regulator $\Theta^3$ at order $\mathcal{O}(\partial^4)$ (see
Table~\ref{table_suppl}). In that case, it makes no sense anymore to
choose as optimal value the extremum among regulators because the
precision of the DE has reached a point where exponents coming from
the various families of regulators spread around the exact value.
When an overlap between two consecutive orders have been reached,
there is no reason to expect that the results represent bounds for a
given quantity and it is necessary to go back to the estimate
$\bar Q^{(s)}$ described in Sect.~\ref{pessimistic}. More generally,
without a strong reason supporting that a certain order of the DE
gives bounds on a certain physical quantity, it is safer to use the
previously presented more conservative central value and error bar. As
an example, for the exponent $\eta$ there is no broad empirical
experience or theoretical information that could lead us to think that
the order $\mathcal{O}(\partial^6)$ gives a bound for the exponent
(except from an extrapolation of the behavior at previous
orders)\footnote{In fact, $\mathcal{O}(\partial^6)$ does not give a
  bound on this exponent but we only know that by exploiting the good
  estimates obtained for this exponent by other means. In any case
  when we have no strong reasons for assuming that results are bounds,
  the more pessimistic estimate should be used.}. At order
$\mathcal{O}(\partial^6)$ a single model has been studied and only two
exponents have been calculated. This does not give us enough
experience to use improved estimates of central values and error bars
but it does give us a good control of the {\it previous} order
$\mathcal{O}(\partial^4)$ that we can exploit when studying the $O(N)$
models.

In the case of $O(N)$ models up to order $\mathcal{O}(\partial^4)$ there are 
strong
indications that the raw results for the considered exponents coming from the DE 
are, in many cases, bounds for the actual values
for $\nu$ and $\eta$ and,
accordingly, we will consider the improved estimate of exponents in those cases. 
This seems clearly to be the case for $\eta$ for all values of $N$ and for $\nu$
for moderate values of $N$, at least for $1\le N\leq 5$. It is important, however, 
to point out that in non-unitary cases ($N=0$
and $N=-2$)
that we analyze in Sect.~\ref{nonunitaryN}, the DE does not seem to 
show consistent bounds on the exponents for
$\eta$ and $\nu$. In the same way, when $N$ is large ($N \gtrsim 10$), there are 
indications that the DE expansion does not give bounds for the exponent $\nu$.
Let us note, however, that the dependence on the regulator becomes very small 
when $N$ grows, making the optimization of
the regulator a much less relevant issue in that limit.

The case of exponent $\omega$ is
different. As explained below the estimates of various orders
of the DE for this exponent 
do not seem to correspond to bounds in any particular domain of $N$. For this
exponent we employ the more conservative estimate 
of central values and
errors presented in Sect.~\ref{pessimistic}. It is interesting to note that even
the concavities
of the curves of $\omega$ as a function of $\alpha$ changes for $N\sim 1$ (see Fig.~\ref{omegaN1vsalpha}
and Figs.~\ref{N2vsalpha} and \ref{N0vsalpha} below).

\subsection{Other sources of error}
\label{err_reg}

In this section we analyze two other sources of error to be taken into account.

First, in the previous analysis, only the error associated with the distance between 
some central value
and the exact value coming from the systematic error of the DE has been 
considered. However, in the cases where the improvement presented
in Sect.~\ref{improverror} can not be considered, one must add a further source 
of uncertainty. That is, when considering
various families of regulators, we made the choice of the center of the interval 
of studied families of regulators but there is no definite reason to 
take one value or the other. As an estimate of error coming
from the uncertainty due to the dependence on the family of regulators, we 
choose for any quantity $Q$ the distance between the two
extreme
values obtained among the families of regulators considered and we call it $\Delta_{reg} Q$.
This source of uncertainty in most studied cases is much smaller than the one coming from 
the difference between one order of the DE and the following.
However, it turns out that at order $\mathcal{O}(\partial^6)$, one can not 
neglect it. In all cases that we choose the
non-optimized central value $\bar Q$ we will choose as error bar $\Delta 
Q=\Delta_{reg}Q+\bar\Delta Q$ (see Table~\ref{N1estimatesanderrors}).

We now analyze a second possible source of error. When the estimates
coming from the DE are not bounds on a given quantity $Q$ it may
happen that the estimates of two consecutive orders of the DE cross
when we vary some parameter, such as the space dimension $d$ or the
number of components of the field $N$ \footnote{This difficulty can
  not take place in the ``improved'' version because in that case,
  successive orders are disjoint.}.  In those cases, the error
estimate presented in Sect.~\ref{pessimistic} is no longer appropriate
because the difference between two consecutive orders is accidentally
small (see Ref.~\cite{DePolsi2019} where the same phenomena takes
place for other exponents).  In that case, it is more convenient to
use a typical value of the error bars and not a particular value which
is near the ``crossing''. This difficulty is encountered in practice
for the exponent $\omega$ at order $\mathcal{O}(\partial^2)$ because
the values for this exponent at that order crosses those of the LPA
for $N$ between 3 and 4 (as explained in detail in
Sect.~\ref{N3and4}). To avoid such difficulty, we exploit the fact
that the exponent $\omega$ becomes exact in the large-$N$ limit. As
such, we impose the error to be, at each order of the DE, a
monotonically decreasing function of $N$. This avoids the artificial
reduction of error bars for $N=3,4$ and 5.  Of course, this can give a
pessimistic error bar but, as stated before, it is preferable to use
conservative error bars than the opposite. A similar difficulty takes
place at order $\mathcal{O}(\partial^4)$ for $\omega$ because the
results from $\mathcal{O}(\partial^2)$ and $\mathcal{O}(\partial^4)$
cross around $N\sim 2$.  In that case, however, one can exploit the
error bars calculated at order $\mathcal{O}(\partial^2)$ to estimate a
very conservative error bars at order $\mathcal{O}(\partial^4)$
without needing to assume a monotonic behavior of error bars. In order
to do so, we adopted the following criterion: whenever the estimated
error for $\omega$ at order $\mathcal{O}(\partial^4)$ is smaller than
the error calculated at order $\mathcal{O}(\partial^2)$ divided by
four, we adopt this last expression. Doing so, we exclude abnormally
small estimates of error bars due to the crossing. In practice, this
augment is advocated when evaluating error estimates for $N=0,1$ and
2.

\section{Critical exponents for $O(N)$ models: Derivative Expansion results}
\label{sec_results}

In the present section we extend previous results to $O(N)$ models at
order $\mathcal{O}(\partial^4)$ of the DE. We first compute the
correction to scaling exponent $\omega$ for $N=1$, which was not
studied previously at this order. We then analyze other values of $N$
and compute the leading exponents $\eta$ and $\nu$ and the correction
to scaling exponent~$\omega$.


Before considering each particular value of $N$,
it is worth mentioning that the nature of error bars are different in
the various studies. CB are able (in many cases) to give rigorous
bounds on the values of critical exponents, under mild assumptions on
the spectrum of operators for unitary theories. When quoting CB
results we employed for all positive values of $N$ such rigorous
bounds when available for exponents $\eta$ and $\nu$. In the case
$N=0$ and for $\omega$ most results in the literature within the CB do
not have the same level of rigor. In those cases, we should keep in
mind that error bars do not have the same meaning as for exponents
$\eta$ and $\nu$ in unitary models.  In MC studies, statistical error
bars are well under control but have a probabilistic
interpretation. Other systematic sources of error are much more
difficult to handle but for the quoted MC studies, they seem to be
under control and consistent with other estimates. For perturbation
theory, high-temperature expansion and DE results, the error bars do
not have the same level of rigor. They depend on assumptions and on
semi-empirical analysis of the results at various orders.
Other estimates of various methods seem to give consistent results but
we observe that some perturbative error bars seem to be too optimistic
because the state-of-the-art results are not within their uncertainty
range.  This is the case, for example, for the recent $\epsilon^6$
results of Ref.~\cite{Kompaniets:2017yct} where the resummation
technique and the methodology to determine error bars is presented in
great detail.  However, some of their results are incompatible with
the most precise results of the literature. The authors of ref
\cite{Kompaniets:2017yct} mention this point but they suggest that it
is too soon to know if the discrepancies of $\epsilon^6$ results with most
precise estimates is significative or not and they suggest to wait to
$\epsilon^7$ results in order to decide.

In the present study, error bars are to be understood as a bracketing
of the exact values. For all critical exponents and values of $N$ that
we have considered, we obtain results that are
compatible, within error bars, with the most precise estimates in the
literature (whenever a prediction more precise than ours is
available). The only exception is the value of $\omega$ for $N=100$ at
order $O(\partial^4)$ of DE. In this case, it may be that we
underestimate the error bars [from our $O(\partial^4)$ result or
from large-$N$ expansion]. Even in that case, error bands almost
overlap.

\subsection{Results for critical exponents for $N=1$}
\label{omegaN1}

\begin{figure}
\includegraphics[width=8cm]{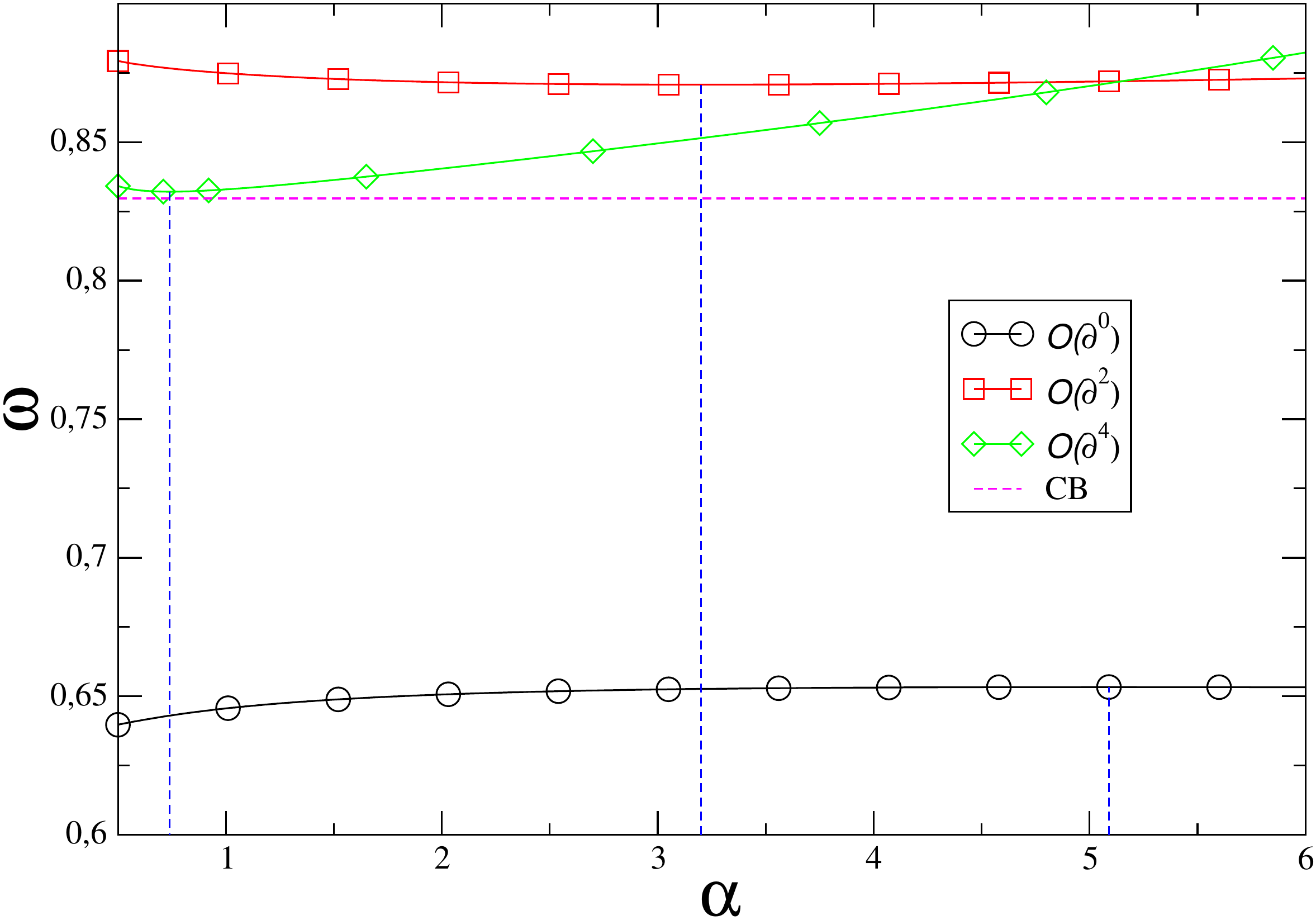}
     \caption{\label{omegaN1vsalpha} Exponent $\omega$ as a function of $\alpha$ for $N=1$ for the
     regulating function (\ref{regulator-exp}).}
\end{figure}

Let us consider first the raw data for the correction to scaling
exponent $\omega$ that can be seen in Table~\ref{table_supplN1}. In
the present work we focus on the regulators that were analyzed in
Ref.~\cite{Balog:2019rrg}, that can be employed at order
$\mathcal{O}(\partial^4)$. When $N=1$, as for the case of $\eta$ and
$\nu$, the raw data for the critical exponent $\omega$ gives PMS
results at successive orders of the DE [up to order
$\mathcal{O}(\partial^4)$] which are disjoints. However, as seen in
Fig.~\ref{omegaN1vsalpha}, both $\mathcal O(\partial^2)$ and
$\mathcal O(\partial^4)$ curves present a minimum, which indicates
that the various orders of the DE are not bounds on this critical
exponent. The same behavior was observed for other regulators.
As a consequence, we use for this exponent the non-improved
estimate of central values and errors presented in
Sect.~\ref{pessimistic}. We observe, nevertheless, a very fast
convergence achieving a precision of the same order of MC estimates,
but as for most other methods in the literature, we have a larger
error bar than for leading exponents $\nu$ and $\eta$.

\begin{table}[]
\caption{\label{table_supplN1} Raw data for $N=1$ of the critical exponent $\omega$ in 
$d=3$ obtained with various families of regulators at
various orders of the DE. The results are also compared to previous results
of the DE. When a value of $\alpha$ different of PMS is employed, this is explicitly indicated.
The results of the CB \cite{Simmons-Duffin:2016wlq} are given 
for comparison.}
\begin{ruledtabular}
\begin{tabular}{lll}
      & \hspace{-0.5cm}regulator           &  $\omega$ \\
\hline
LPA   &$W$                                 &  0.6541         \\
      &$\Theta^1$ \cite{Litim02}           &  0.6557       \\
      &$\Theta^3$                          &  0.6551       \\
      &$E$                                 &  0.6533        \\           
      &Power-law \cite{Morris:1997xj}      &  0.63        \\
      &$S$ \cite{Hasenfratz86}             & 0.595                 \\
      \hline   
$O(\partial^2)$
     &$W$                                  &  0.8702       \\
     &$\Theta^3$                           &  0.8698     \\
     &$E$                                  &  0.8707      \\
     &Power-law \cite{Morris:1997xj}       &  0.897        \\
     \hline
$O(\partial^4)$ 
     &$W$                                  &  0.8313       \\
     &$\Theta^3$                           &  0.8310     \\
     &$E$                                  &  0.8321      \\
      \hline
CB \cite{Simmons-Duffin:2016wlq}  &  &  0.82968(23)  
\end{tabular}
\end{ruledtabular}
\end{table}

Before considering other values of $N$ let us sum up the results
obtained up to now for the three dimensional Ising universality class,
presented in Table~\ref{expN1final}.  It is worth mentioning that the
results are very precise (particularly for $\nu$ and $\omega$). At
first sight one could get the impression that the order
$\mathcal{O}(\partial^6)$ does not improve the results significantly
with respect to order $\mathcal{O}(\partial^4)$ for $\eta$ and
$\nu$. However, this only reflects our poorer experience on the
behavior of the DE at order $\mathcal{O}(\partial^6)$ and the
consequent use of a much more pessimistic estimate of central values
and error bars. In fact, by looking directly at the raw data presented
in Table~\ref{table_suppl} one observes that the DE does give better
estimates for any regulator at successive orders, including order
$\mathcal{O}(\partial^6)$.

Another strategy in order to estimate central values followed in
Ref.~\cite{Balog:2019rrg} is to exploit the whole series of data for a
given exponent in order to extrapolate the central value and error
bars. This strategy gives better estimates of central values and a
smaller error bar. However, we follow here a strategy that can be
implemented for $O(N)$ models where we only have at our disposal the
results for the DE up to order $\mathcal{O}(\partial^4)$. More
generally, we propose a general method that can be employed safely for
very general models where, in most cases, the DE has only been
studiedup to order $\mathcal{O}(\partial^2)$.

\begin{table}[]
\caption{\label{expN1final} Final results 
at various orders of the DE with appropriate error bars for
$N=1$ in $d=3$. 
Results for $\eta$ and $\nu$ are taken from \cite{Balog:2019rrg}.
Results of the CB (\cite{Kos:2014bka} for $\eta$ and $\nu$ and 
\cite{Simmons-Duffin:2016wlq} for $\omega$),
MC \cite{Hasenbusch10}, High-temperature expansion 
\cite{Campostrini:2002cf}, 
and 6-loop, $d=3$ perturbative RG values \cite{Guida:1998bx}, and 
$\epsilon-$expansion at order $\epsilon^5$ \cite{Guida:1998bx}
and at order $\epsilon^6$ \cite{Kompaniets:2017yct}
are also given for comparison.}
\begin{ruledtabular}
\begin{tabular}{lllll}
                   &  $\nu$         &  $\eta$      &  $\omega$ \\
                   \hline
LPA                & 0.64956        &  0           &  0.654   \\
$O(\partial^2)$    & 0.6308(27)     &  0.0387(55)  &  0.870(55)   \\
$O(\partial^4)$    &  0.62989(25)   &  0.0362(12)  &  0.832(14)    \\
$O(\partial^6)$    &  0.63012(16)   &  0.0361(11) &      \\
      \hline
CB      &   0.629971(4)  & 0.0362978(20) & 0.82968(23)\\
6-loop,  $d=3$    &   0.6304(13)   &  0.0335(25)  & 0.799(11) \\
$\epsilon-$expansion, $\epsilon^5$     & 0.6290(25) & 0.0360(50) & 0.814(18)\\
$\epsilon-$expansion, $\epsilon^6$     & 0.6292(5)   & 0.0362(6) & 0.820(7)\\
High-T.          &  0.63012(16)  &  0.03639(15)  & 0.83(5) \\
MC           &   0.63002(10)  &  0.03627(10)  & 0.832(6)
\end{tabular}
\end{ruledtabular}
\end{table}

\subsection{The controversial $N=2$ case: the Derivative Expansion take}

The $N=2$ case describes the important $XY$ universality class that
corresponds to many physical systems, including easy planes magnetic
systems and the $\lambda$-transition of the Helium-4 superfluid.  For
a classical review of various systems in this universality class, we
refer to \cite{Pelissetto02}.  The $O(2)$ case is particularly
important because, as discussed in the Introduction, there is a
longstanding controversy concerning the value of the critical exponent
$\nu$ between the most precise experiments\footnote{Indeed, the
  critical exponent that is actually measured is the specific heat
  exponent $\alpha$ for the transition of the superfluid helium 4, that
  can be related to $\nu$ by a hyper-scaling relation.} \cite{He4exp}
and the best theoretical estimates given by some MC simulations
\cite{Campostrini_2006,Xu:2019mvy} and very recent CB results
\cite{chester2019carving}. Most field-theoretical methods
\cite{Guida:1998bx,Kos:2016ysd} (including CB before
\cite{chester2019carving}) have been unable to decide the point
because of the high level of precision reached by experiments and
simulations.  Indeed, as discussed in \cite{Xu:2019mvy}, there is even
a discrepancy among various MC results that in some cases give results
compatible with experiments \cite{lan2012highprecision}, but a
consensus seems to have been reached that the most precise simulations
\cite{Campostrini_2006,Xu:2019mvy,Hasenbusch:2019jkj} are very far
away from the experimental prediction.  We present now our
$O(\partial^4)$ DE estimate of critical exponents $\eta$, $\nu$ and
$\omega$.

\begin{figure}
\includegraphics[width=8cm]{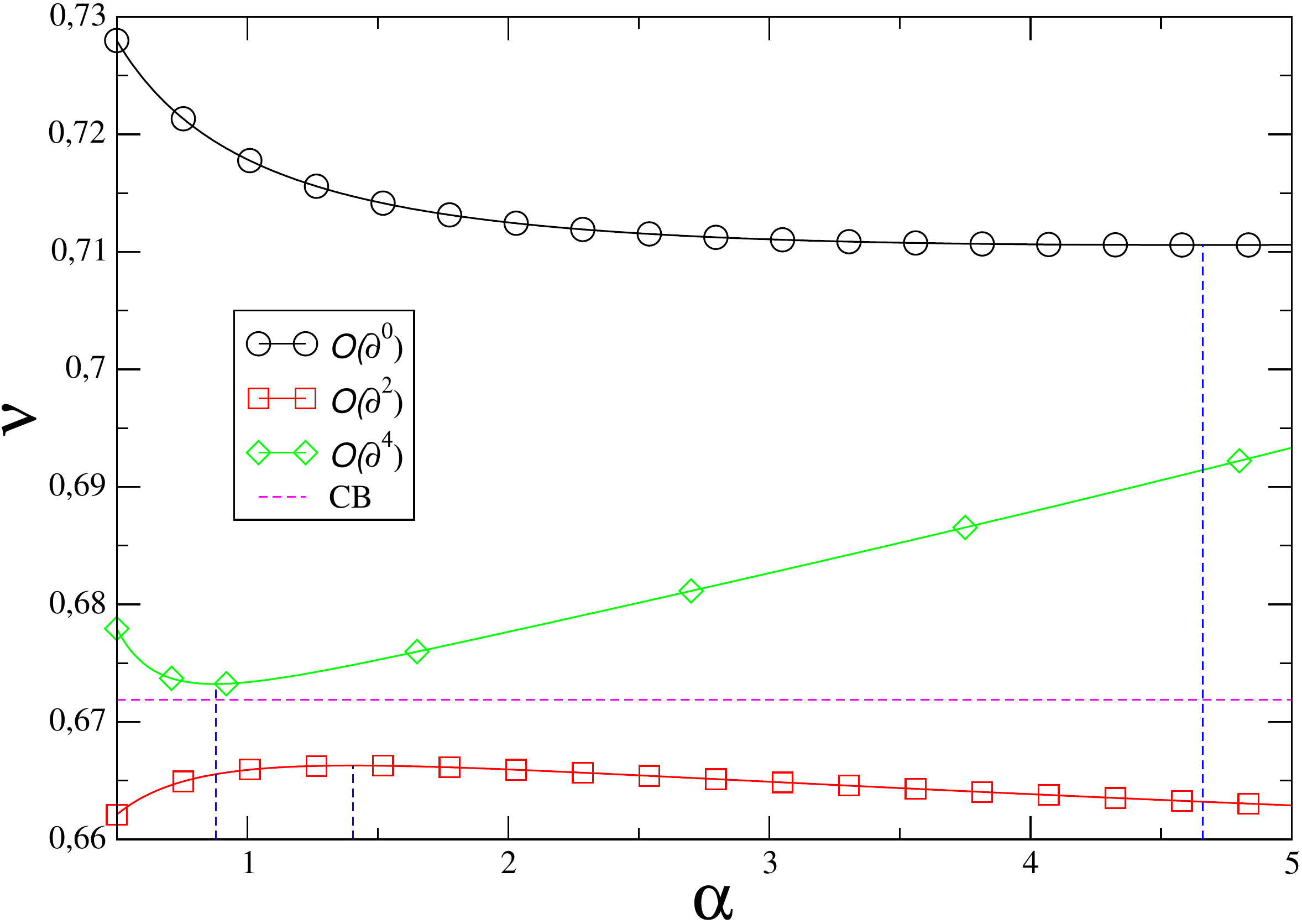}\\
\includegraphics[width=8cm]{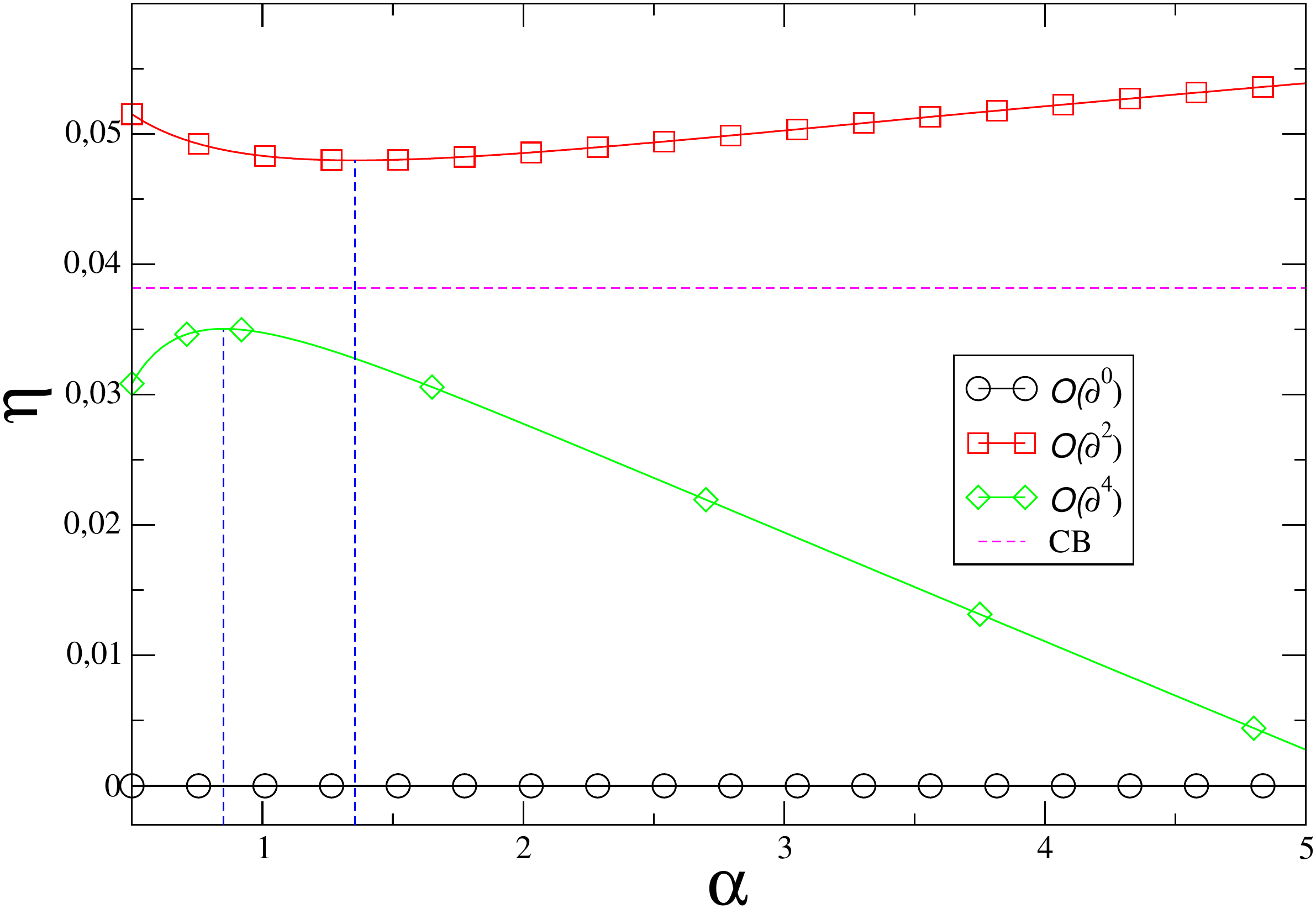}\\
\includegraphics[width=8cm]{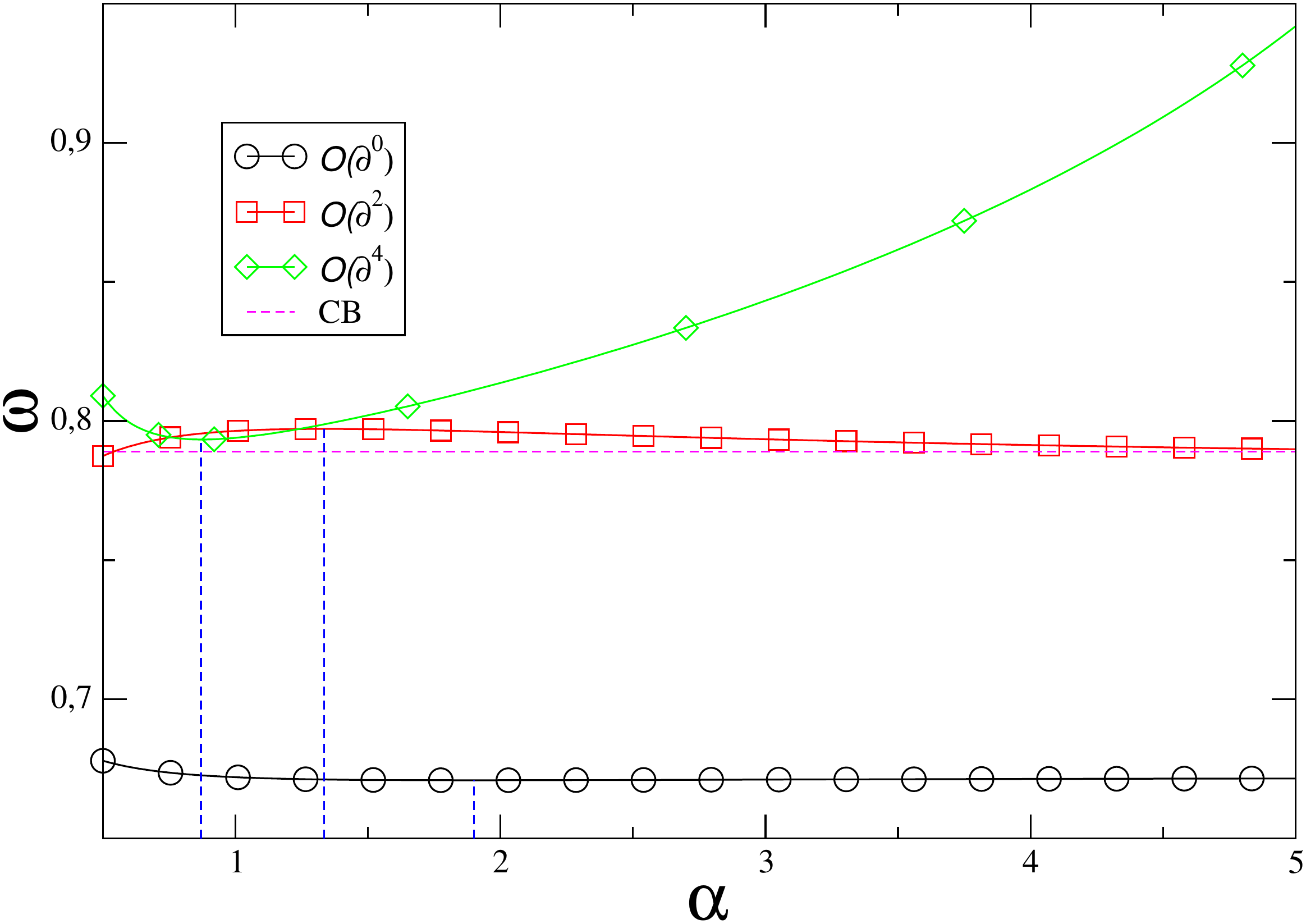}
     \caption{\label{N2vsalpha} From top to bottom: exponents $\nu$, $\eta$ and $\omega$ as a
     function of $\alpha$ for $N=2$ for the
     regulating function (\ref{regulator-exp}).}
\end{figure}

The raw data for these exponents obtained at successive orders of the
DE for the same regulators mentioned in previous section are presented
in Table~\ref{table_supplN2} in Appendix~\ref{apprawdata}.  We also
included in this table the previous results obtained with the DE.  As
for $N=1$, for all considered families of regulators the concavity of
the curves of exponents $\eta$ and $\nu$ as a function of the
parameter $\alpha$ alternates, see Fig.~\ref{N2vsalpha}. Moreover, the
results at successive orders of the DE are disjoint, which strongly
indicates alternating bounds on the critical exponents at this order
of the DE. Accordingly, we employ the improved estimate of central
values and error bars presented in Sect.~\ref{improverror} for those
exponents. The corresponding results are presented in
Table~\ref{expN2final} where they are compared to other results in the
literature both theoretical and experimental. Special attention must
be given to the exponent $\omega$ where it is seen in
Fig.~\ref{N2vsalpha} that the results at order $\mathcal(\partial^2)$
and $\mathcal(\partial^4)$ intersect. Moreover, the LPA curve, which
is below the $\mathcal O(\partial^2)$ one, presents a minimum, not a
maximum.  The various orders of the DE definitely do not give bounds
on that exponent. We therefore use for this exponent the more
conservative estimate of error bars described in
Sect.~\ref{pessimistic}. In what concerns $\omega$, we shall use the
same pessimistic error bar for other values of $N$

We reach for the three exponents very precise estimates. In
particular, we obtain a better precision than perturbative
estimates. We do not reach, however the level of accuracy of MC
\cite{Campostrini_2006,Xu:2019mvy,Hasenbusch:2019jkj} and (for $\nu$
and $\eta$) from the CB \cite{chester2019carving} which appeared
during the completion of this work.  We find, for the controversial
value of $\nu$, a result that turns out to be compatible with the most
precise MC simulations and CB and incompatible with
experiments from \cite{He4exp}.

MC and CB clearly give more precise determinations of the critical
exponents. We would like to mention, however, that the numerical
effort is much bigger in these two methods than the one we had to
face. Typically, finding a fixed point takes about 2 hours in a laptop
while MC involved several years of CPU time and CB
about $10^2$ years ofCPU time.

\begin{table}[]
\caption{\label{expN2final} Final results 
at various orders of the DE with appropriate error bars for
$N=2$ in $d=3$. 
Results to the CB from 2016 (\cite{Kos:2016ysd}
for $\eta$ and $\nu$ and \cite{Echeverri2016} for $\omega$)
and also from 2019 \cite{chester2019carving}, combined MC and High-Temperature analysis from \cite{Campostrini_2006} and recent (2019) MC from \cite{Hasenbusch:2019jkj}, 
and 6-loop, $d=3$ perturbative RG values \cite{Guida:1998bx}, and 
$\epsilon-$expansion at order $\epsilon^5$ \cite{Guida:1998bx} and order $\epsilon^6$ \cite{Kompaniets:2017yct}
are also given for comparison. Results for most precise experiments are also included: Helium-4 superfluid from \cite{He4exp} and \cite{PhysRevB.30.5103} for $\nu$, XY-antiferromagnets (CsMnF$_3$ from \cite{Oleaga_2014} and SmMnO$_3$ from \cite{PhysRevB.85.184425}), and XY-ferromagnets
(Gd$_2$IFe$_2$ and Gd$_2$ICo$_2$ from \cite{REISSER1995265}). Whenever needed, scaling relations are used in order
to express results in terms of $\eta$ and $\nu$.}
\begin{ruledtabular}
\begin{tabular}{llll}
                 &   $\nu$         &  $\eta$    &  $\omega$ \\
\hline
LPA              &   0.7090        &  0          &  0.672         \\
$O(\partial^2)$  &   0.6725(52)    &  0.0410(59) &  0.798(34)       \\
$O(\partial^4)$  &   0.6716(6)     &  0.0380(13) &  0.791(8)    \\
      \hline
CB (2016) & 0.6719(12)   &  0.0385(7)        & 0.811(19)    \\
CB (2019) & 0.6718(1)    &  0.03818(4)    & 0.794(8)    \\
6-loop $d=3$     & 0.6703(15)	  & 0.0354(25) & 0.789(11) \\
$\epsilon-$expansion, $\epsilon^5$     & 0.6680(35) & 0.0380(50) &	0.802(18) \\
$\epsilon-$expansion, $\epsilon^6$     & 0.6690(10) & 0.0380(6) &	0.804(3) \\
MC+High-T. (2006) & 0.6717(1)      &  0.0381(2)	   & 0.785(20)\\
MC (2019) & 0.67169(7)      &  0.03810(8)	   & 0.789(4)\\
\hline
Helium-4 (2003)     & 0.6709(1)	& & \\
Helium-4 (1984)     & 0.6717(4) & & \\
XY-AF (CsMnF$_3$)  & 0.6710(7) \\
XY-AF (SmMnO$_3$)  & 0.6710(3) \\
XY-F (Gd$_2$IFe$_2$) & 0.671(24) & 0.034(47)\\
XY-F (Gd$_2$ICo$_2$) & 0.668(24) & 0.032(47)\\
\end{tabular}
\end{ruledtabular}
\end{table}

\subsection{Results for some physically interesting cases}
\label{N3and4}
We present now result for two other physically relevant cases with
positive integer values of $N$ (and, as such, unitary). These are the
Heisenberg universality class $N=3$, relevant for isotropic
ferromagnets, and the $N=4$ universality class relevant for the chiral
phase transition in the physics of strong interactions. We refer to
\cite{Pelissetto02} for a detailed description of various systems in
these two universality classes.

We present now our results at successive orders of the DE up to order
$\mathcal{O}(\partial^4)$ for critical exponents $\eta$, $\nu$ and
$\omega$. The raw data for these exponents obtained at successive
orders of the DE (for the same regulators mentioned in previous
sections) are presented in Table~\ref{table_supplN3} and
\ref{table_supplN4} in Appendix~\ref{apprawdata}.  For completeness,
results from previous DE analysis are also included in these tables.
The same considerations as for $N=1$ and $N=2$ applies here,
concerning the strong indication that successive orders of the DE give
bounds on exponents $\eta$ and $\nu$ but not for $\omega$. As such, we
implement the ``improved'' version of central values and error
estimates for the first two exponents but not for $\omega$.

\begin{figure}
     \begin{picture}(216,165)
      \put(0,0){\includegraphics[width=216pt]{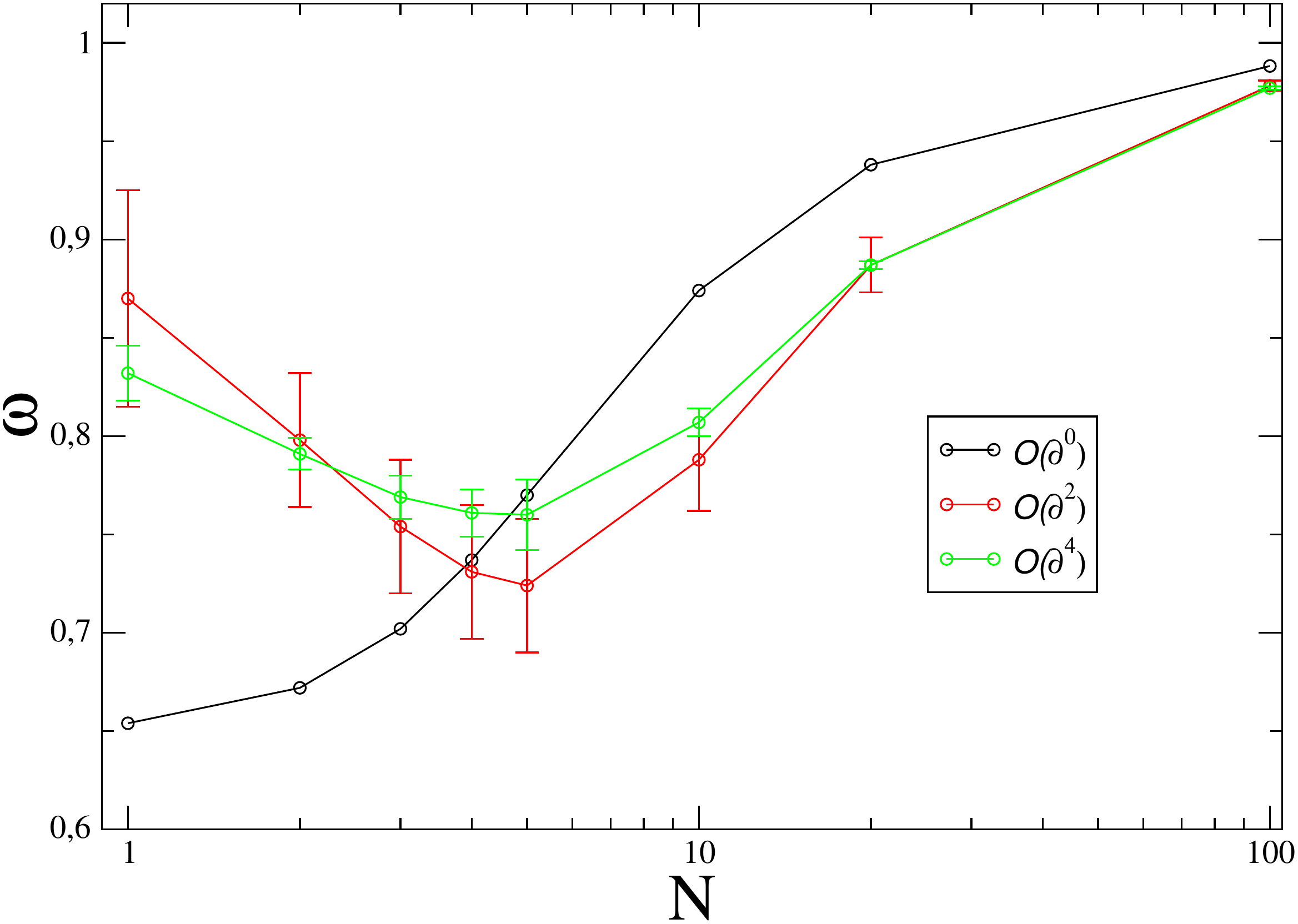}}
     \end{picture}
     \label{omegavsN}
     \caption{Exponent $\omega$ as a function of $1/N$ for $N\ge 1$ at various orders of the 
DE.}

\end{figure}

A special mention must be made for the calculation of error bars for
the exponent $\omega$. We employ the most conservative estimates for
this exponent and a particular analysis (already discussed in
Sect.~\ref{err_reg}) must be done in that case.  Indeed, as shown in
Fig.~\ref{omegavsN}, when varying $N$ the exponent $\omega$ turns out
to be relatively stable at orders $O(\partial^2)$ and $O(\partial^4)$
but varies in a very important way at order LPA. More importantly, the
curve of orders LPA and $O(\partial^2)$ crosses in a point between
$N=3$ and $N=4$.  Similarly, the curves for orders $O(\partial^2)$
and $O(\partial^4)$ cross in a point for $N\sim 2$.  These
exceptional points where two consecutive orders of the DE cross can
make the uncertainty presented in Sect.~\ref{errorbars} too
optimistic.  To avoid this artificially small error bar, we employ a
conservative estimate of error bars explained in Sect.~\ref{err_reg}.
The corresponding results are presented in Tables~\ref{expN3final} and
\ref{expN4final} where, as before, they are compared to other results
in the literature both theoretical and experimental.

We obtain again very precise estimates for the three exponents that,
in some cases, are the most precise exponents obtained in the
literature for these universality classes from field-theoretical
methods. The results are in some cases even more precise than MC
simulations.  Moreover, all our results are compatible (within error
bars) with the best estimates in the literature (whenever more precise
results than ours are available). This is a strong indication that our
estimates of error bars are reliable. In Table~\ref{expN3final}
experimental results are also presented for various physical
realizations of $N=3$ universality class. As for $N=2$, the
experimental precision for the exponent $\eta$ is much lower than for
exponent $\nu$.

\begin{table}[]
\caption{\label{expN3final} Final results 
at various orders of the DE with appropriate error bars for
$N=3$ in $d=3$. 
For reference results of CB (\cite{Kos:2016ysd}
for $\eta$ and $\nu$ and \cite{Echeverri2016} for $\omega$), MC 
(\cite{PhysRevB.84.125136} for $\eta$ and $\nu$ and \cite{Hasenbusch_2001} for 
$\omega$), 
combined MC and High-Temperature analysis from \cite{Campostrini_2002}, 
and 6-loop, $d=3$ perturbative RG values \cite{Guida:1998bx}, and 
$\epsilon-$expansion at order $\epsilon^5$ \cite{Guida:1998bx} and order $\epsilon^6$ \cite{Kompaniets:2017yct}
are also given for comparison.  Results for most precise experiments are also included
(Isotropic ferromagnets Gd$_2$BrC and Gd$_2$IC from \cite{PhysRevB.52.3546} and CdCr$_2$Se$_4$ from \cite{refId0}).
 Whenever needed, scaling relations are used in order
to express results in terms of $\eta$ and $\nu$.}
\begin{ruledtabular}
\begin{tabular}{llll}
                 &   $\nu$         &  $\eta$    &  $\omega$ \\
\hline
LPA              &   0.7620        &  0          &  0.702        \\
$O(\partial^2)$  &   0.7125(71)    &  0.0408(58) &  0.754(34)       \\
$O(\partial^4)$  &   0.7114(9)     &  0.0376(13) &  0.769(11)    \\
      \hline
CB               & 0.7120(23)     & 0.0385(13) &  0.791(22)   \\
6-loop $d=3$     & 0.7073(35)	& 0.0355(25) & 0.782(13) \\
$\epsilon-$expansion, $\epsilon^5$     & 0.7045(55)   & 0.0375(45) & 0.794(18) \\
$\epsilon-$expansion, $\epsilon^6$     & 0.7059(20)   & 0.0378(5)  & 0.795(7) \\
MC               &  0.7116(10)    & 0.0378(3) & 0.773 \\
MC+High-T.       &  0.7112(5)     & 0.0375(5) & \\
\hline
Ferromagnet Gd$_2$BrC  & 0.7073(43) & 0.032(10) \\
Ferromagnet Gd$_2$IC & 0.7067(60) & 0.061(15) \\
Ferromagnet CdCr$_2$Se$_4$ & 0.656(56) & 0.041(23)
\end{tabular}
\end{ruledtabular}
\end{table}

\begin{table}[]
\caption{\label{expN4final} Final results 
at various orders of the DE with appropriate error bars for
$N=4$ in $d=3$. 
For reference results of CB ($\eta$ and $\nu$ from \cite{Kos:2015mba} and $\omega$ from \cite{Echeverri2016}), MC ($\eta$ and $\nu$ from \cite{PhysRevE.73.056116} and $\omega$ from \cite{Hasenbusch_2001}), 
and 6-loop, $d=3$ perturbative RG values \cite{Guida:1998bx} and
$\epsilon-$expansion at order $\epsilon^5$ \cite{Guida:1998bx} and order $\epsilon^6$ \cite{Kompaniets:2017yct} and
 are also given for comparison.  }
\begin{ruledtabular}
\begin{tabular}{llll}
                 &   $\nu$         &  $\eta$    &  $\omega$ \\
\hline
LPA              &   0.805        &  0          &  0.737       \\
$O(\partial^2)$  &   0.749(8)     &  0.0389(56) &  0.731(34)      \\
$O(\partial^4)$  &   0.7478(9)    &  0.0360(12) &  0.761(12)  \\
      \hline
CB               &  0.7472(87)    &  0.0378(32) &  0.817(30)   \\
6-loop $d=3$     & 0.741(6)	& 0.0350(45) & 0.774(20) \\
$\epsilon-$expansion, $\epsilon^5$     & 0.737(8) & 0.036(4) & 0.795(30) \\
$\epsilon-$expansion, $\epsilon^6$     & 0.7397(35) & 0.0366(4) & 0.794(9) \\
MC                &  0.7477(8)     & 0.0360(4) & 0.765
\end{tabular}
\end{ruledtabular}
\end{table}

\subsection{The large $N$ case}
\label{largeNsect}

Even if the $N=5$ has been proposed to describe a possible universality
class in some superconductors \cite{Pelissetto02}, the main purpose of
the present section is to test our DE results in a limit where
different kinds of approximations have been implemented, including the
Large-$N$ expansion. In fact, the expressions for the critical
exponents $\eta$, $\nu$ and $\omega$, in this limit have been computed
at next-to-next-to-leading order~
\cite{Okabe78,Vasil'ev1982,Broadhurst:1996ur} :
\begin{align}
\eta&=\frac{8}{3\pi^2}\frac 1 N 
-\frac{512}{27\pi^4}\frac{1}{N^2}-\frac{8}{27\pi^6 N^3} \nonumber\\
 &\times\Big[\frac{797}{18}-\zeta(2)\Big(27\log(2)-\frac{61}{4}\Big)
 +\zeta(3)\frac{189}{4}\Big]
+\mathcal{O}\big(1/N^4\big)\nonumber\\
 \nu&=1-\frac{32}{3\pi^2}\frac 1 N -\frac{128}{27 \pi^4}
 \big(-112+27\pi^2\big)\frac{1}{N^2}+\mathcal{O}\big(1/N^3\big) \nonumber\\
\omega&=1-\frac{64}{3\pi^2}\frac{1}{N}+\frac{128}{9\pi^4}\frac{1}{N^2}
\Big(\frac{104}{3}-\frac{9\pi^2}{2}\Big)+\mathcal{O}\big(1/N^3\big)
\end{align}
We used these expressions as reference values. As well known, the large-$N$ 
expansion is expected to be a good approximation only for $N$ larger than about ten. However, for reference, we
compare also to this expansion in the $N=5$ case. In order to estimate central 
values and error bars of the $1/N$ expansion, we use a very conservative 
estimate:
we choose as central value the exponent obtained at the 
highest known order in the $1/N$ expansion and we estimate the error bar as the difference 
between this order
and the previous one. This estimate may be too pessimistic for $N$ 
large
enough and the actual errors bars at $N=20$ or 100 could be 
smaller. However, given that coefficients in the large-$N$ expansion 
are typically not of order one, we employed this conservative estimate. It is important to note that even with 
such conservative error bars some results of the large-$N$ expansion becomes 
incompatible with other estimates for $N=5$ and 10. For some of the considered values of $N$ there are also 
available resummed 6-loops and MC results that we include for comparison. 

\begin{figure}
\includegraphics[width=8cm]{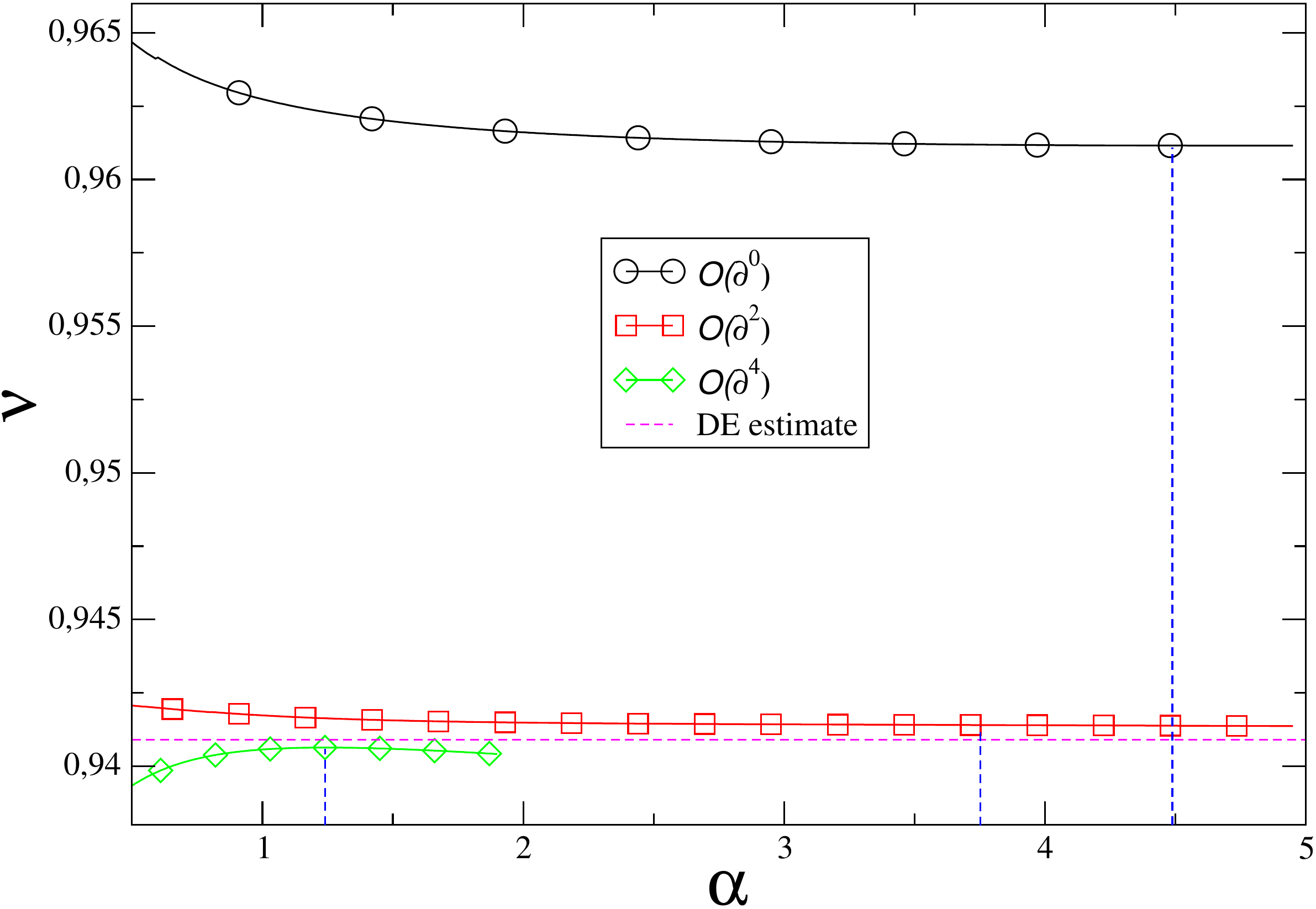}\\
\includegraphics[width=8cm]{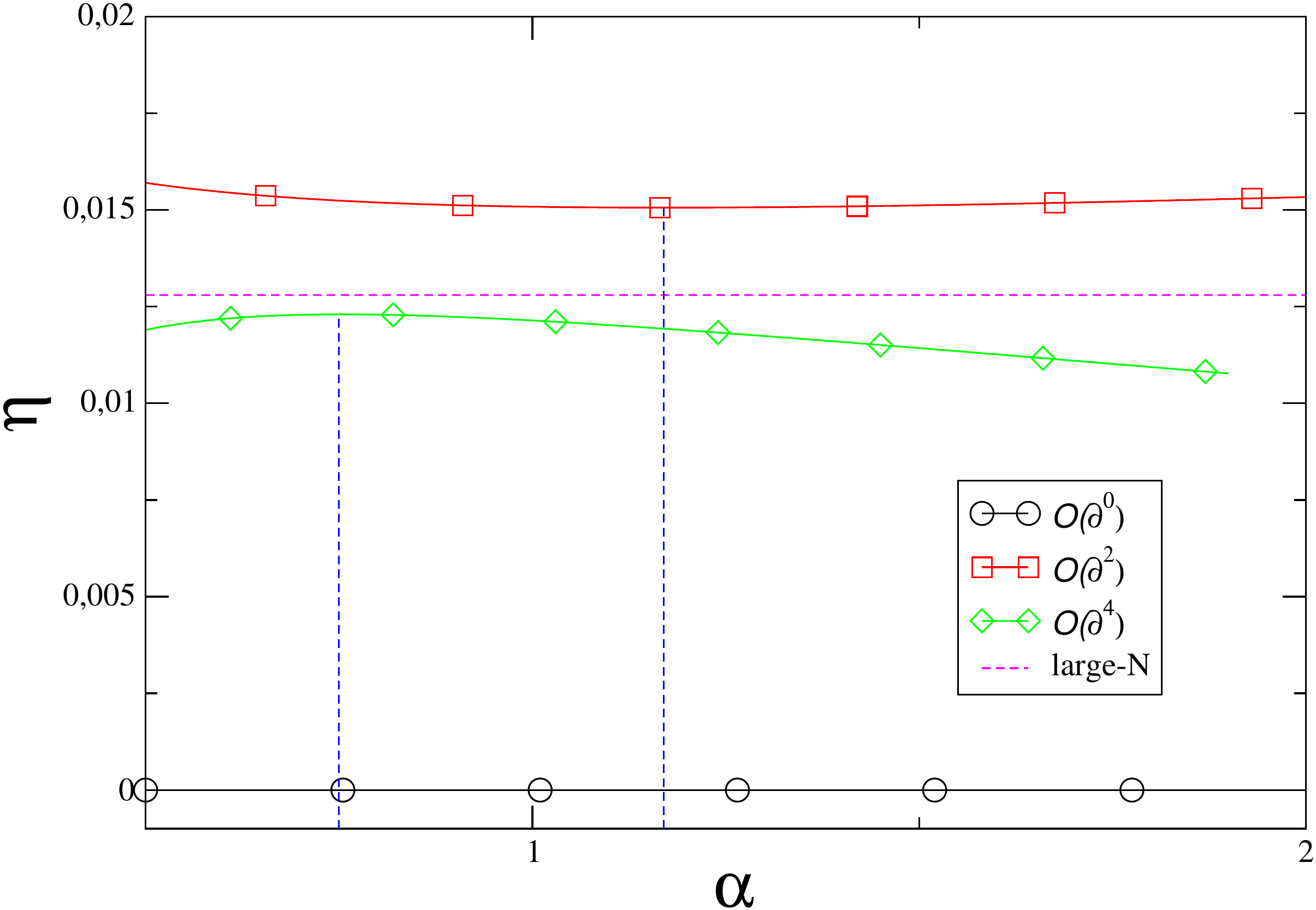}\\
\includegraphics[width=8cm]{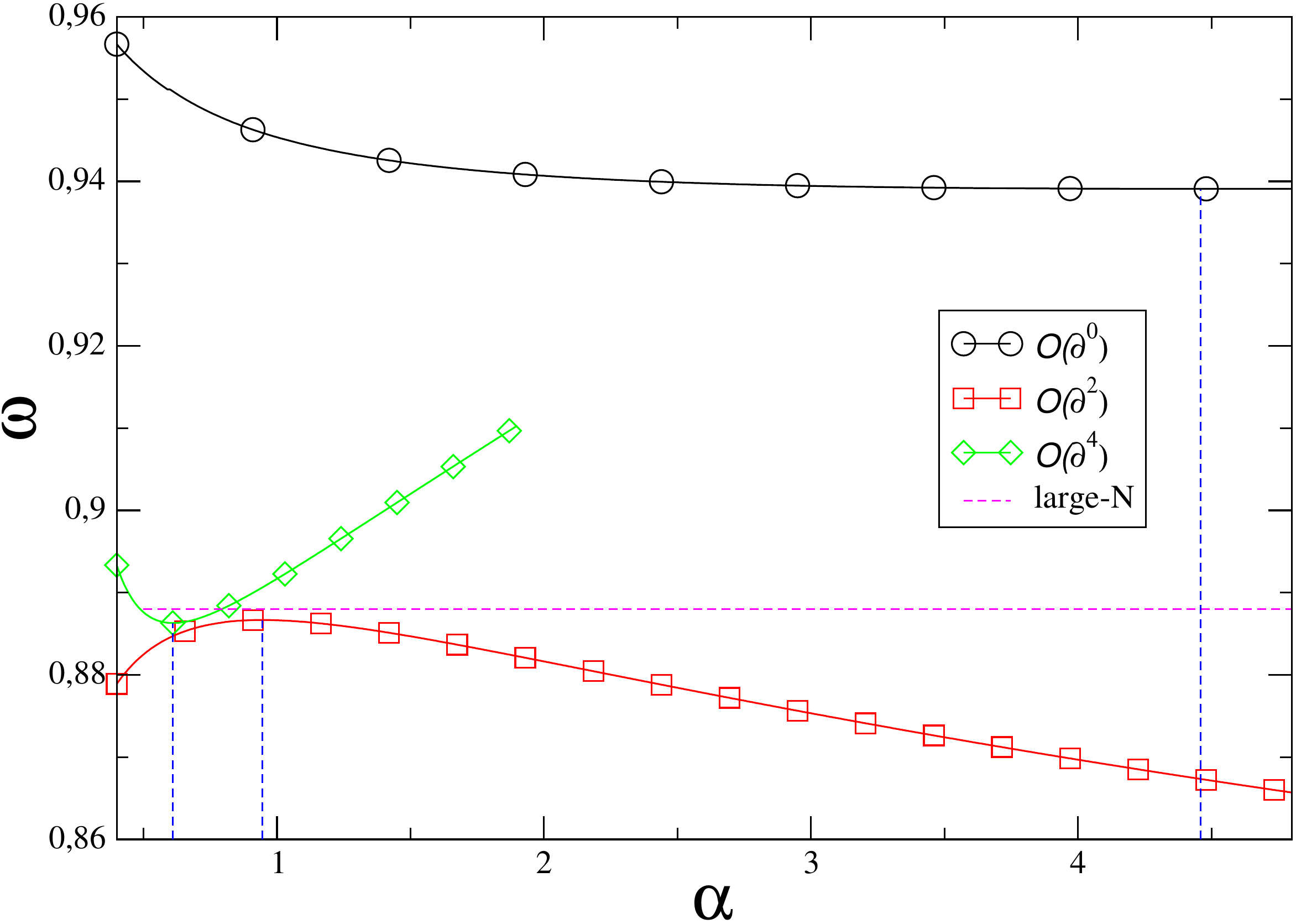}\\
     \caption{\label{N20vsalpha} From top to bottom: exponents $\nu$, $\eta$ and $\omega$ as a
     function of $\alpha$ for $N=20$ for the
     regulating function (\ref{regulator-exp}).}
\end{figure}

In order to estimate our central values and error bars we employed for
all the considered values of $N> 5$ and for the exponents $\nu$ and
$\omega$ the most conservative estimate presented in
Sects.~\ref{pessimistic} and \ref{err_reg}. The reason is that we do
not have clear indications that for those values of $N$ the estimates
coming from the DE constitute bounds on those critical exponents. Even
more, in some cases for $N\ge 10$ we observe overlaps between the
values obtained in consecutive orders of the DE for these exponents,
indicating that, at least for those orders and values of
$O(\partial^4)$, the hypothesis of being bounds is not fulfilled, see
Fig.~\ref{N20vsalpha}.  The case of the exponent $\eta$ is different
because we observe the same qualitative behavior for all $N\ge 1$ which
strongly indicates that, at least up to order $O(\partial^4)$, the
estimates are bounds on the exponents (as a typical example, see
Fig.~\ref{N20vsalpha}).  In any case, the dependence on regulator
families and on the regulating function parameter $\alpha$ becomes
much less pronounced for $N$ large enough.  Accordingly, the relevance
of the precise choice of regulator becomes less important.  Our
results turn out to be mostly compatible with other estimates in the
literature and seem to be even more precise.  Indeed, only for very
large values of $N$ -- of order 20 -- the $1/N$ expansion becomes more
precise than our $\mathcal O(\partial^4)$ results. Of course, as is well known,
the large $N$ limit is obtained exactly \cite{DAttanasio:1997yph} with
the DE already at order LPA, but we also observe that $1/N$
corrections seem to be very well estimated at order $\mathcal O(\partial^4)$.

\begin{table}[]
\caption{\label{expN5final} Final results 
at various orders of the DE with appropriate error bars for
$N=5$ in $d=3$. 
For reference results of MC \cite{Hu_2001}, Large-$N$ expansion 
\cite{Okabe78,Vasil'ev1982,Broadhurst:1996ur}
and 6-loop, $d=3$ perturbative RG values \cite{Antonenko_1995}
are also given for comparison.}
\begin{ruledtabular}
\begin{tabular}{llll}
                 &   $\nu$         &  $\eta$    &  $\omega$ \\
\hline
LPA              &   0.839        &  0          &  0.770       \\
$O(\partial^2)$  &   0.782(8)     &  0.0364(52) &  0.724(34)      \\
$O(\partial^4)$  &   0.7797(9)    &  0.0338(11) &  0.760(18)  \\
      \hline
6-loop $d=3$     &   0.766  & 0.034 &     \\
MC               &  0.728(18)   &  &  \\
Large-$N$        &  0.71(7)  & 0.031(15) & 0.51(6)
\end{tabular}
\end{ruledtabular}
\end{table}

\begin{table}[]
\caption{\label{expN10final} Final results 
at various orders of the DE with appropriate error bars for
$N=10$ in $d=3$. 
For reference from Large-$N$ expansion 
\cite{Okabe78,Vasil'ev1982,Broadhurst:1996ur}
and 6-loop, $d=3$ perturbative RG values \cite{Antonenko_1995}
are also given for comparison. }
\begin{ruledtabular}
\begin{tabular}{llll}
                 &   $\nu$         &  $\eta$    &  $\omega$ \\
\hline
LPA              &   0.919        &  0          &  0.874       \\
$O(\partial^2)$  &   0.877(11)    &  0.0240(34) &  0.788(26)      \\
$O(\partial^4)$  &   0.8776(10)   &  0.0231(6)   &  0.807(7)  \\
      \hline
6-loop $d=3$     &   0.859   & 0.024 &     \\
Large-$N$        &  0.87(2)  & 0.023(2) & 0.77(1)
\end{tabular}
\end{ruledtabular}
\end{table}

\begin{table}[]
\caption{\label{expN20final} Final results 
at various orders of the DE with appropriate error bars for
$N=20$ in $d=3$. 
For reference results of CB \cite{Kos:2015mba}, Large-$N$ expansion 
\cite{Okabe78,Vasil'ev1982,Broadhurst:1996ur}
and 6-loop, $d=3$ perturbative RG values \cite{Antonenko_1995}
are also given for comparison.}
\begin{ruledtabular}
\begin{tabular}{llll}
                 &   $\nu$         &  $\eta$    &  $\omega$ \\
\hline
LPA              &   0.9610        &  0          &  0.938       \\
$O(\partial^2)$  &   0.9414(49)    &  0.0130(19) &  0.887(14)      \\
$O(\partial^4)$  &   0.9409(6)    &  0.0129(3)  &   0.887(2)  \\
      \hline
CB               &   0.9416(87)   &  0.0128(16) &\\      
6-loop $d=3$     &   0.930        & 0.014 &     \\
Large-$N$        &  0.941(5)      & 0.0128(2) & 0.888(3)
\end{tabular}
\end{ruledtabular}
\end{table}

\begin{table}[]
\caption{\label{expN100final} Final results 
at various orders of the DE with appropriate error bars for
$N=100$ in $d=3$. 
For reference results from Large-$N$ expansion 
\cite{Okabe78,Vasil'ev1982,Broadhurst:1996ur}
is also given for comparison. }
\begin{ruledtabular}
\begin{tabular}{llll}
                 &   $\nu$         &  $\eta$      &  $\omega$ \\
\hline
LPA              &   0.9925        &  0           &  0.9882       \\
$O(\partial^2)$  &   0.9892(11)    &  0.00257(37)  &  0.9782(26)      \\
$O(\partial^4)$  &   0.9888(2)     &  0.00268(4)&  0.9770(8)  \\
      \hline
Large-$N$        &   0.9890(2)     & 0.002681(1)  & 0.9782(2)
\end{tabular}
\end{ruledtabular}
\end{table}

\subsection{Analysis of some non-unitary cases: $N=0$ and $N=-2$}
\label{nonunitaryN}

In this section we consider two cases of $O(N)$ models for non-positive
values of $N$.  These are interesting for two different
reasons. First, they describe situations of physical interest.  $N=0$
corresponds to self-avoiding walks \cite{degennes72} which model long
polymer chain with self-repulsion.  The case $N=-2$ corresponds to
loop-erased random walks \cite{lawler1980}. In such a random walk
every loop is erased when it is formed. Second, these cases are
interesting because unitarity is probably not valid when $N$ is not a
positive integer. Indeed, for positive integer values of $N$ and $d$,
$O(N)$ models have a clear interpretation in terms of a
Ginzburg-Landau field theory verifying reflection-positivity. However,
in the cases that are obtained by analytical continuation, the
validity of unitarity of the Minkowskian version of the model is far
from obvious. Since unitarity was explicitly used in the proof of the
convergence of DE, we have to analyze the convergence properties in
this situation. In this sense, the cases $N=0$ and $N=-2$ can be seen
as benchmarks for non-unitary theories. A similar issue occurs in the
case of analytical continuation to non-integer $d$. It has been
pointed out that unitarity is lost \cite{Hogervorst:2015akt} in this
situation, which makes the CB program more difficult to implement (at
least with the same level of rigor as for positive integer values of
$d$).

\begin{figure}
\includegraphics[width=8cm]{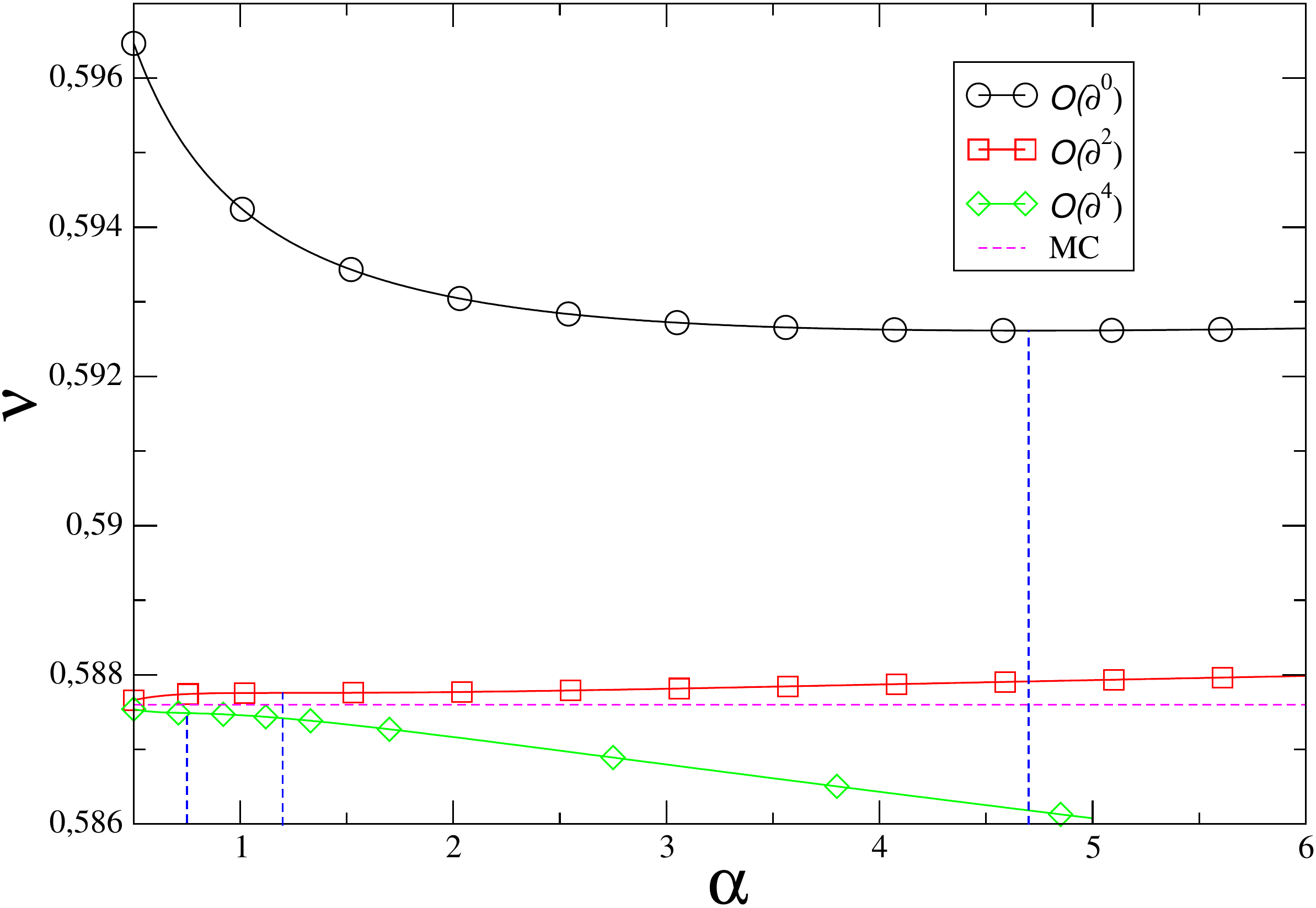}\\
\includegraphics[width=8cm]{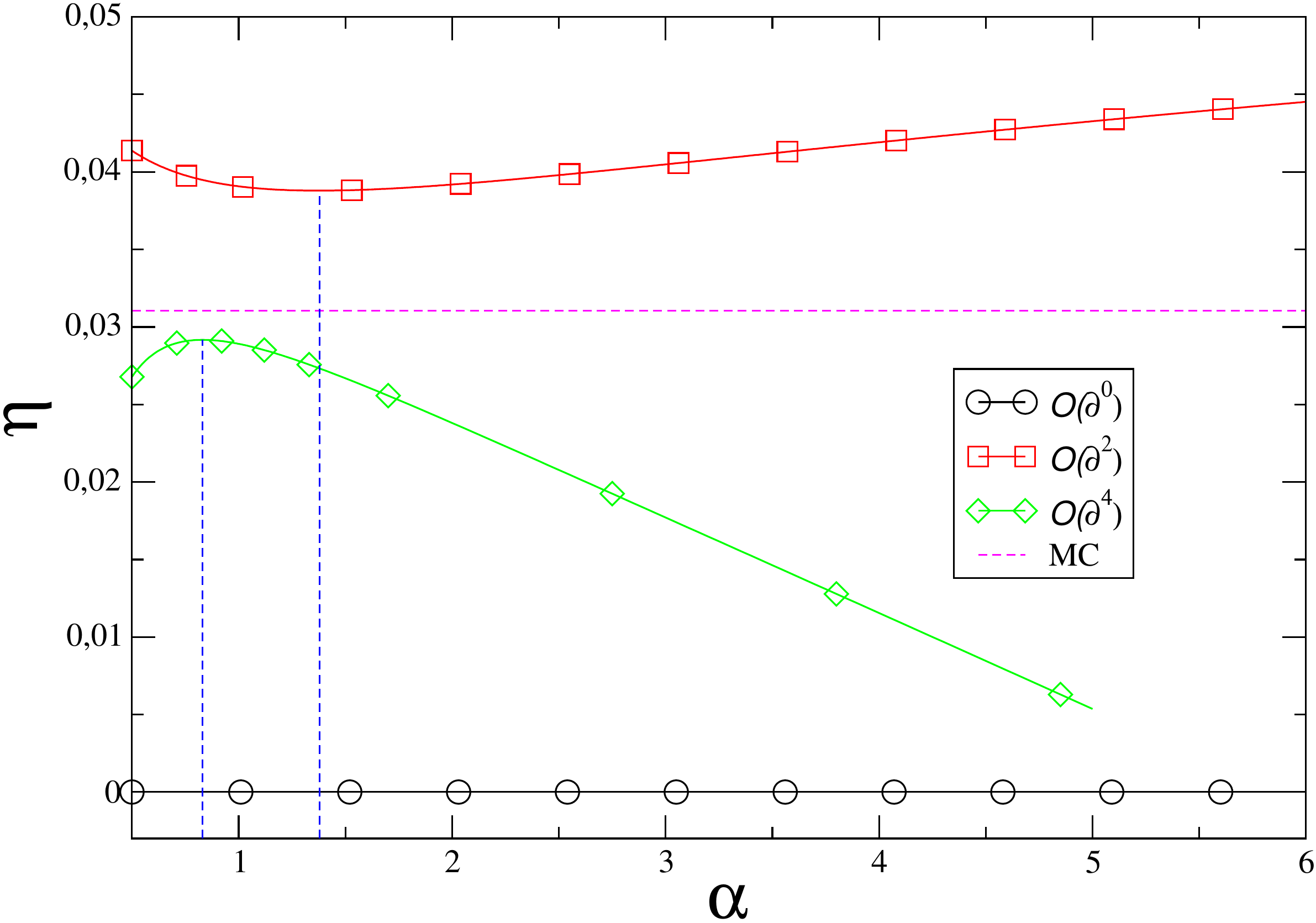}\\
\includegraphics[width=8cm]{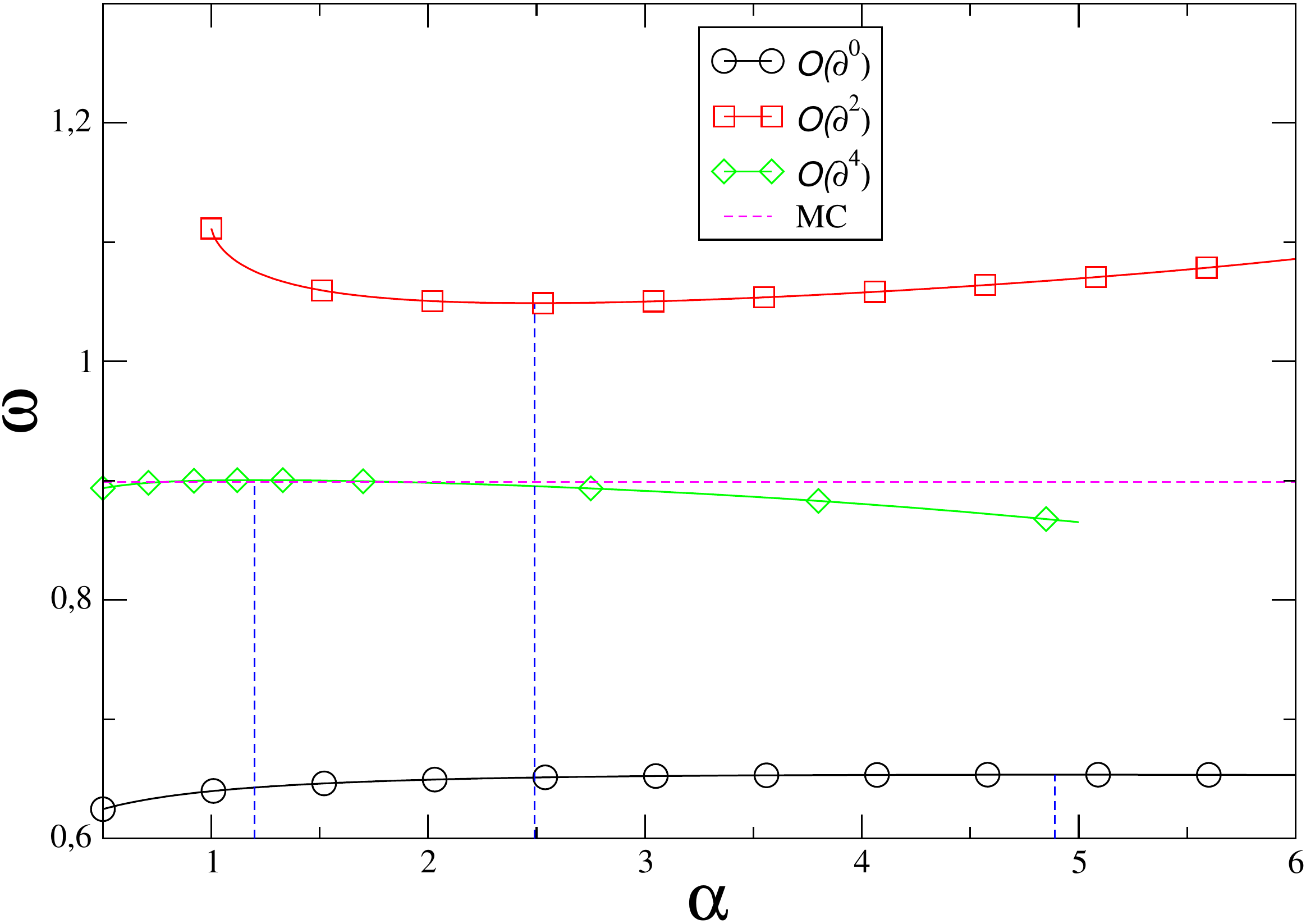}
     \caption{\label{N0vsalpha} From top to bottom: exponents $\nu$, $\eta$ and $\omega$ as a
     function of $\alpha$ for $N=0$ for the
     regulating function (\ref{regulator-exp}).}
\end{figure}

Let us mention, however, that the estimates on the convergence of the
DE from Ref.~\cite{Balog:2019rrg} do not rely on all the information
coming from the structure of a unitary theory but only on the position
of singularities on the complex plane of squared momenta.  In
particular, at all orders of perturbation theory, these singularities
are located, for all values of $N$ including negative values, at the
same positions as for unitary theories. As a consequence, at least at
all orders of perturbation theory, one should expect that our estimate
of the convergence of the DE and the existence of a small parameter
should remain correct. Of course, the information coming from
unitarity is that this structure remains correct
non-perturbatively. Another information that comes from unitarity is
that series for correlation functions are alternating (at least at
large order). This comes from the fact that it is dominated by the 2
or 3-particles threshold which has a definite sign due to
unitarity. For $N$ that are not positive integers there is no reason
at all to believe that successive orders of the DE give bounds on
exponents. For example, see Fig.~\ref{N0vsalpha} where there is no
indication of alternating values for $\nu$ (but the results for $\eta$
does seem to alternate).

From a practical point of view we will continue to assume that there
is a relative factor of order 4 that suppresses successive orders of
the DE in order to estimate error bars but we will employ the more
conservative estimate that do not assume that they give bounds on the
exponents.  The raw data obtained at various orders with various
regulators are presented in Tables~\ref{table_supplN0} and
\ref{table_supplN-2}. They indicate that these assumptions seem to be
justified.  The results with corresponding error bars are presented in
Tables~\ref{expN0final} and \ref{expN-2final} which seem to confirm
that our methodology for estimating errors (at least the most
conservative version) remains valid for those non-positive values of
$N$. It must be pointed out that the results of the CB for $N=0$ are
not as rigorous as for positive integer values of $N$. As a
consequence, in this cases their error bars does not constitute
rigorous error bars. In Table~\ref{expN0final} we also compare to an
experimental realization of the $N=0$ case in a polymer solution
\cite{refId1}.

A special mention must be done for some exact results known for the
$N=-2$ case.  In fact, the exponents $\eta$ and $\nu$ are known
exactly to take their mean-field value 0 and 1/2.  This result is
well-known from an all-order perturbative analysis (see, for example,
\cite{ZinnJustin:2002ru}) but a nice proof going beyond perturbation
theory has been proposed recently
\cite{Wiese:2018dow,Wiese:2019xmu}. In these references, the exponent
$\omega$ (which is not known exactly) is also estimated to be
$\omega=0.83\pm 0.01$. It is interesting that the DE recovers these
results with high precision. Indeed, it was observed a long time ago
that the LPA order reproduces the exact result for $\eta$ and $\nu$
exactly for $N=-2$ \cite{Morris:1997xj}. However, the reason that
makes the result exact at LPA order is somehow too simple. The fact,
that $\eta=0$ at LPA order is by construction true for any
$N$. Therefore, its coincidence with the exact value for $N=-2$ is
then an accident. Moreover, the flow of the mass term at zero field in
LPA is controlled exclusively by the 4-point vertex at zero momenta
and field which, for any $N$, verifies
\begin{equation}
 \Gamma^{(4)}_{ijkl}(p_i=0;\phi=0)\propto 
\big(\delta_{ij}\delta_{kl}+\delta_{ik}\delta_{jl}+\delta_{il}\delta_{jk}\big).
\end{equation}
Now, the flow of the mass at zero field is proportional to
\begin{equation}
 \Gamma^{(4)}_{ijkk}(p_i=0;\phi=0)\propto \delta_{ij}\big(N+2\big).
\end{equation}
The consequence is that the mass parameter does not flow for $N=-2$ and then 
$\nu=1/2$.

This simple analysis does not extend beyond LPA and the exactness of $\eta=0$ 
and $\nu=1/2$ is slightly
spoiled by the DE approximation. In order to remain true, very non-trivial 
relations must be preserved along
the flow that, given the analysis of \cite{Wiese:2018dow,Wiese:2019xmu} must be 
exact but are only satisfied
within DE (beyond LPA) approximately. The systematic errors seem to grow in
the raw data when going from order $O(\partial^2)$ to order $O(\partial^4)$
as seen in Table~\ref{table_supplN-2}. Let us mention, however, that the relations $\eta=0$ 
and $\nu=1/2$ are extremely well satisfied and
the exact values are obtained in all cases within the expected error bars
as seen in Table~\ref{expN-2final} \footnote{It is interesting to observe that, apparently,
the distance with exact results is as expected at order $O(\partial^4)$ but is abnormally small
at order $O(\partial^2)$ (see Table~\ref{table_supplN-2}). We do not have an explanation for
such high precision at order $O(\partial^2)$.}. One must point out that the estimate of 
systematic error bars is, however, problematic for those exponents at $N=-2$. As 
mentioned in Sect.~\ref{err_reg}, when two consecutive orders
of the DE cross, our estimates of error bars are not justified and it is better 
to employ a ``typical'' value
of error bars. In the present analysis we employ as ``typical'' value for these 
two exponents those of $N=0$.
For the non-trivial exponent $\omega$ one can employ without difficulty our error
bar estimate and results are
compatible, and with the same order of precision as, the one coming from perturbation theory.

The study of these two non-unitary models then suggests that the
domain of application of our methodology for estimating error bars
goes beyond the realm of unitary theories.

\begin{table}[]
\caption{\label{expN0final} Final results 
at various orders of the DE with appropriate error bars for
$N=0$ in $d=3$. 
For reference results of CB \cite{Shimada:2015gda}, MC \cite{Clisby16,Clisby_2017}, 
Length doubling method series \cite{Schram_2017}, 
and 6-loop, $d=3$ perturbative RG values \cite{Guida:1998bx}, and 
$\epsilon-$expansion at order $\epsilon^5$ \cite{Guida:1998bx} and order $\epsilon^6$ \cite{Kompaniets:2017yct}
are also given for comparison. Results for most precise experiment are also included (polystyrene benzene dilute solutions \cite{refId1}. Whenever needed, scaling relations are used in order
to express results in terms of $\eta$ and $\nu$.}
\begin{ruledtabular}
\begin{tabular}{llll}
                 &   $\nu$         &  $\eta$      &  $\omega$ \\
\hline
LPA              &   0.5925        &  0           &  0.66       \\
$O(\partial^2)$  &   0.5879(13)    &  0.0326(47)  &  1.00(19)      \\
$O(\partial^4)$  &   0.5876(2)     &  0.0312(9)   &  0.901(24)  \\
      \hline
CB      &  0.5876(12)   &  0.0282(4) &     \\
Series LDM       & 0.58785(40) & 0.0327(22) &\\
MC      & 0.58759700(40) &  0.0310434(30)  & 0.899(14)   \\
6-loop $d=3$     & 	0.5882(11)	& 0.0284(25) & 0.812(16) \\
$\epsilon-$expansion, $\epsilon^5$     & 	0.5875(25) & 0.0300(50) &	0.828(23)\\
$\epsilon-$expansion, $\epsilon^6$     & 	0.5874(3)  & 0.0310(7) &	0.841(13)\\
\hline
Polymer solution & 0.586(4)
\end{tabular}
\end{ruledtabular}
\end{table}

\begin{table}[]
\caption{\label{expN-2final}  Final results 
at various orders of the DE with appropriate error bars for
$N=-2$ in $d=3$. 
For reference results from exact or perturbative results \cite{Wiese:2018dow,Wiese:2019xmu}.}
\begin{ruledtabular}
\begin{tabular}{llll}
                 &   $\nu$         &  $\eta$      &  $\omega$ \\
\hline
LPA              &   1/2        &  0              &  0.700       \\
$O(\partial^2)$  &   0.5000(12) &  0.0000(47)     &  0.84(19)      \\
$O(\partial^4)$  &   0.5001(1)  &  0.0004(9)      &  0.838(24)  \\
      \hline
exact/ 6-loop     &  1/2        &  0              &   0.83(1)    \\
\end{tabular}
\end{ruledtabular}
\end{table}

\section{Conclusion and perspectives}

The DE of the NPRG equations has proved to be an approximation 
scheme which is versatile and capable of tackling a very broad range of 
physical systems. However, until very recently, the success of such a method
remained suspicious because of the apparent lack of a control parameter in order to 
estimate {\it a priori} the precision
of the results. In a recent work \cite{Balog:2019rrg} this important difficulty 
has been addressed and it was
shown that the DE has a control parameter of order $1/9$ to $1/4$. This has been shown in two ways. First,
on a theoretical basis, by considering the position of the singularities on the 
complex plane of 1PI correlation functions as a function of momenta. 
Second, by corroborating this general analysis with an empirical study of the 
precision of the DE at large order [$\mathcal{O}(\partial^6)$] 
when applied to the critical regime of a Ginzburg-Landau model in the Ising 
universality class.

In the present article we used this general analysis to study the
critical exponents $\eta$, $\nu$ and $\omega$ of an important family
of critical phenomena characterized by $O(N)$-invariant
Ginzburg-Landau models.  Previous studies performed within the DE at
order $\mathcal{O}(\partial^2)$ had shown good precision but in the
present work we show that when going to order
$\mathcal{O}(\partial^4)$ one achieves, in most cases, the best
precision for those systems with field-theoretical methods
\footnote{The only exceptions are the $N=1$
  \cite{ElShowk:2012ht,El-Showk:2014dwa,Kos:2014bka} and $N=2$ cases
  \cite{chester2019carving} in which quasi-exact results of the
  CB are available and for very large values of $N$
  ($N\gtrsim 20$) where the large-$N$ expansion becomes very precise
  \cite{Okabe78,Vasil'ev1982,Broadhurst:1996ur}.}.  In some cases we
were even able to attain a better precision than Monte-Carlo
estimates.  In order to perform this analysis we developed a
systematic procedure to compute error bars within the DE. This
procedure was corroborated by a careful analysis of the very precisely
studied Ising universality class (corresponding to $N=1$) obtained in
Ref.~\cite{Balog:2019rrg} and extending it to various values of $N$ in
the three-dimensional case.

An important application is the analysis of the $O(2)$ or XY model
where a longstanding controversy exists between experiments
\cite{He4exp} and the state-of-the-art Monte-Carlo estimates
\cite{Campostrini_2006,Xu:2019mvy,Hasenbusch:2019jkj} and very recent
results from the CB \cite{chester2019carving} for the
specific heat $\alpha$ (or, equivalently, the correlation length
exponent, $\nu$). Most theoretical estimates, based on fixed dimension
re-summed perturbation theory, $\epsilon-$expansion or previous
CB works were unable to achieve a precision high
enough to disentangle between the estimates of experiments and
simulations. Our results are in agreement with Monte-Carlo simulations
and new CB results but exclude the results
obtained with Helium-4 in micro-gravity. This result can be interpreted
in many ways. One possible explanation is the one proposed in
\cite{Kos:2016ysd}: It could be that the analysis of the experimental
data made in Ref.~\cite{He4exp} underestimate error
bars. Alternatively we could consider other possible sources of
systematic errors in the experiment whose exceptional realization in
micro-gravity makes difficult to repeat. Another possible explanation,
but much more challenging from the theoretical viewpoint could be that
for some unexplained reason the $O(2)$ model does not describe
properly the Helium-4 critical point. This is difficult to believe
because scale-invariant theories are typically a discrete set and
there is no doubt that the $O(2)$ model describes at least three
digits of critical exponents of Helium-4 superfluid transition. This
explanation would require another scale-invariant model extremely
close to but different from the $O(2)$ model. In any case, the
agreement between two independent theoretical estimates pushes in
favour of a new realization of the experiment in order to confirm or
discard previous experimental results.

The results of the present article paves the way towards many
applications in the near future. First, the methodology used to
estimate error bars within the DE can be applied in many applications
within NPRG [even at order $\mathcal{O}(\partial^2)$]. Second, this
analysis of error estimates, even if probably very pessimistically,
applies also to other approximation schemes such as the
Blaizot-M\'endez-Wschebor
scheme~\cite{Blaizot:2005xy,Benitez09,Benitez:2011xx}. This
possibility should be exploited because many finite momentum physical
properties are beyond the reach of the DE.  Third, in the present
article we only considered the two independent dominant exponents and
the correction to scaling exponent of $O(N)$ universality class at
order $\mathcal{O}(\partial^4)$. It is clear that with the same
methodology we can analyze a very broad set of universal and
non-universal properties of $O(N)$ models well studied in the
literature with other methods (see, for example, \cite{Pelissetto02}
for many universal aspects that could be analyzed with the present
setup).  Given the precision reached for leading critical exponents is
to be expected that we can improve for several quantities the best
current theoretical estimates by using the DE at order
$\mathcal{O}(\partial^4)$.  Fourth, on more fundamental aspects, the
present analysis is strongly based on the use of ``Principle of
Minimal Sensitivity'' that turned out to improve significantly the
results of the DE but requires a more solid theoretical basis. In this
sense, an important under-exploited information that could bring some
clarity on this point may come from the use of conformal symmetry
that, up to now, has almost not been exploited in the NPRG context in
order to improve physical predictions (see, however,
\cite{Rosten:2014oja,delamotte2016scale,Rosten:2016zap,Morris:2018zgy,DePolsi2019}).

\acknowledgments

This work was supported by Grant 412FQ293 of the CSIC (UdelaR)
Commission and Programa de Desarrollo de las Ciencias B\'asicas
(PEDECIBA), Uruguay and ECOS Sud U17E01.  IB acknowledges the support
of the Croatian Science Foundation Project IP-2016-6-7258 and the
QuantiXLie Centre of Excellence, a project cofinanced by the Croatian
Government and European Union through the European Regional
Development Fund - the Competitiveness and Cohesion Operational
Programme (Grant KK.01.1.1.01.0004). NW also thanks the LPTMC for its
hospitality and the CNRS for funding during the autumn of 2019. The
authors would like to thank B. Delamotte for a detailed reading of a
previous version of this manuscript.

\appendix

\section{Raw data for DE estimates of critical exponents}
\label{apprawdata}

\begin{table}
\caption{\label{table_supplN-2} Raw data for $N=-2$ critical exponents in $d=3$ 
obtained with various regulators at various orders of the DE.}
\begin{ruledtabular}
\begin{tabular}{lllll}
      & \hspace{-0.5cm}regulator          &  $\nu$         &  $\eta$ &  $\omega$ 
\\
\hline
LPA   &$W$  				  & 1/2	   &  0      &  0.7000   \\
      &$\Theta^3$	   	          & 1/2	   &  0      &  0.7021    \\
      &$E$ 			          & 1/2	   &  0      &  0.6983 \\           
      &Power-law \cite{Morris:1997xj}     & 1/2    &  0      &    \\ 
      \hline   
$O(\partial^2)$
     &$W$  				  & $0.5+2.8\times 10^{-8}$   &  $5.9\times 10^{-8}$ &  0.8451    \\
     &$\Theta^3$			  & $0.5+5.9\times 10^{-7}$   &  $1.2\times 10^{-6}$ &  0.8447    \\
     &$E$  				  & $0.5+7.4\times 10^{-8}$   &  $1.3\times 10^{-7}$ &  0.8446	   \\
\hline
$O(\partial^4)$
      &$W$			          & $0.5+7.0\times 10^{-5}$   & $8.5\times 10^{-5}$  &  0.8368   \\
      &$\Theta^3$ 			  & $0.5+5.9\times 10^{-5}$   & $9.7\times 10^{-5}$  &  0.8344   \\
      &$E$	 		          & $0.5+8.5\times 10^{-5}$   & $9.2\times 10^{-4}$  &  0.8411   \\
\end{tabular}
\end{ruledtabular}
\end{table}


\begin{table}
\caption{\label{table_supplN0} Raw data for $N=0$ critical exponents in $d=3$ 
obtained with various regulators at various orders of the DE. When a value of $\alpha$ different of
PMS is employed, this is explicitly indicated.}
\begin{ruledtabular}
\begin{tabular}{lllll}
      & \hspace{-0.5cm}regulator          &  $\nu$         &  $\eta$ &  $\omega$ 
\\
\hline
LPA   &$W$  				  & 0.5925	   &  0      &  0.6549   \\
      &$\Theta^1$ \cite{Litim02}	  & 0.5921	   &  0      &  0.6579   \\
 &$\Theta^3$	   	          & 0.5923	   &  0      &  0.6567    \\
      &$E$ 			          & 0.5926	   &  0      &  0.6535	 \\ 
      &Power-law \cite{Morris:1997xj}     & 0.596           &  0      &  0.62  \\ 
\hline   
$O(\partial^2)$
     &$W$  			      & 0.5878 *     &  0.0384 &  1.0407  \\
     &$W$ ($\alpha=1$) \cite{VonGersdorff:2000kp} & 0.590     &  0.039 &    \\
     &$\Theta^3$		& 0.5879 	   &  0.0373 &  0.9431	 \\
     &$E$  				 & 0.5878 *	   &  0.0388 &  1.0489	 \\
\hline
$O(\partial^4)$
      &$W$			          &  0.5875 *	   &  0.0299  & 0.9006   \\
      &$\Theta^3$ 			  &  0.5876	*   &  0.0303  &  0.9007     \\
      &$E$	 		          &  0.5875 * 	   & 0.0292  &  0.9005   \\
\end{tabular}
\end{ruledtabular}
\end{table}


\begin{table}[h!]
\caption{\label{table_supplN2}  Raw data for $N=2$ critical exponents in $d=3$ obtained with various regulators at various orders of the DE.
When a value of $\alpha$ different of PMS is employed, this is explicitly indicated.}
\begin{ruledtabular}
\begin{tabular}{lllll}
      & \hspace{-0.5cm}regulator          &  $\nu$         &  $\eta$ &  $\omega$ \\
\hline
LPA   &$W$                                & 0.7099         &  0      &  0.6717         \\
      &$\Theta^1$ \cite{Litim02}          & 0.7082         &  0      &  0.6712        \\
      &$\Theta^3$                         & 0.7090         &  0      &  0.6715        \\
      &$E$                                & 0.7106         &  0      &  0.6716        \\           
      &Power-law \cite{Morris:1997xj}     & 0.73           &  0      &  0.66  \\
      \hline   
$O(\partial^2)$
     &$W$                                 & 0.6669         &  0.0474 &  0.7983       \\
      &$W$ ($\alpha=1$) \cite{VonGersdorff:2000kp} & 0.666     &  0.049 &    \\
     &$\Theta^3$                          & 0.6673         &  0.0469 &  0.7992     \\
     &$E$                                 & 0.6663         &  0.0480 &  0.7972      \\
     &Power-law \cite{Morris:1997xj}      & 0.65           &  0.044  &  0.38  \\
     \hline
$O(\partial^4)$
      &$W$                                &  0.6725        &  0.0361 &  0.7906    \\
      &$\Theta^3$                         &  0.6722        &  0.0367 &  0.7893     \\
      &$E$                                &  0.6732        &  0.0350 &  0.7934      \\
\end{tabular}
\end{ruledtabular}
\end{table}

In this Appendix we present the raw data for exponents $\nu$, $\eta$
and $\omega$ obtained for various values of $N$ with the regulators
presented in Eqs.~(\ref{regulators}) (in the case of $\Theta^n$
regulators we present in all cases the results for $n=3$ since for
this value of $n$ the DE is well behaved until order
$\mathcal{O}(\partial^4)$ and, for $N=1$, it turned out to be optimum
at that order). For almost all cases, the results are presented at a
PMS, determined as an extremum of the corresponding exponent as a
function of $\alpha$ for each regulator.  The only exceptions are
marked with an asterisk in the tables. In those cases, there are no
PMS for some particular exponents. In order to choose a particular
value of $\alpha$ when no standard PMS is present, we extend the
philosophy of PMS which requires the ``minimum sensitivity''. When no
extremum is present, we verified in each case that the point with
lower sensitivity, in the studied exponents to the parameter $\alpha$
corresponds to an inflexion point and, accordingly, we choose that
value. We also included in the various tables previous DE results when
available.

\begin{table}
\caption{\label{table_supplN3} Raw data for $N=3$ critical exponents in $d=3$ 
obtained with various regulators at various orders of the DE.
When a value of $\alpha$ different of PMS is employed, this is explicitly indicated.}
\begin{ruledtabular}
\begin{tabular}{lllll}
      & \hspace{-0.5cm}regulator          &  $\nu$         &  $\eta$ &  $\omega$ 
\\
\hline
LPA   &$W$                                & 0.7631	   &  0      &  0.7019   \\
      &$\Theta^1$ \cite{Litim02}           & 0.7611	   &  0      &  0.6998   \\
      &$\Theta^3$           		  & 0.7620	   &  0      &  0.7010   
\\
      &$E$   	                          & 0.7639	   &  0      &  0.7026   
\\
      &Power-law \cite{Morris:1997xj}     & 0.78           &  0      &  0.71  \\
   \hline   
$O(\partial^2)$
     &$W$       		                  & 0.7047     &  0.0471 &  0.7541   \\
    &$W$ ($\alpha=1$) \cite{VonGersdorff:2000kp} & 0.704     &  0.049 &    \\
     &$\Theta^3$    		              & 0.7054	   &  0.0466 &	0.7563    \\
     &$E$                                 & 0.7039     &  0.0476 &	0.7516   \\
     &Power-law \cite{Morris:1997xj}     & 0.745           &  0.035      &  0.33  \\
     \hline
$O(\partial^4)$
      &$W$  	                          &  0.7126	   & 0.0358  &  0.7681   
\\
      &$\Theta^3$                         &  0.7122	   & 0.0363  &  0.7659   
\\
      &$E$                        	  &  0.7136	   &  0.0347 &  0.7729   
 \\
\end{tabular}
\end{ruledtabular}
\end{table}

\begin{table}
\caption{\label{table_supplN4} Raw data for $N=4$ critical exponents in $d=3$ 
obtained with various regulators at various orders of the DE.
When a value of $\alpha$ different of PMS is employed, this is explicitly indicated.}
\begin{ruledtabular}
\begin{tabular}{lllll}
      & \hspace{-0.5cm}regulator          &  $\nu$         &  $\eta$ &  $\omega$ 
\\
\hline
LPA   &$W$  				  & 0.8063	   &  0      &  0.7370    \\
      &$W$ ($\alpha=1$) \cite{VonGersdorff:2000kp} & 0.739     &  0.047 &    \\
      &$\Theta^1$ \cite{Litim02}          & 0.8043	   &  0      &  0.7338   \\
      &$\Theta^3$                         & 0.8052	   &  0      &  0.7354    \\
      &$E$                                & 0.8071	   &  0      &  0.7383	 \\
      &Power-law \cite{Morris:1997xj}     & 0.824          &  0      &  0.75  \\
      \hline   
$O(\partial^2)$
     &$W$        	                  & 0.7405     &  0.0450 &  0.7310    \\
     &$\Theta^3$                      & 0.7412	   &  0.0445 &  0.7340   \\
     &$E$      		    		      & 0.7396	   &  0.0455 &  0.7274 	   \\
     &Power-law \cite{Morris:1997xj}     & 0.816          &  0.022      &  0.42  \\
     \hline
$O(\partial^4)$
      &$W$  		          &  0.7490	   & 0.0343  &  0.7588   \\
      &$\Theta^3$           	          &  0.7487	   & 0.0348  &  0.7561   \\
      &$E$                  	          &  0.7500	   & 0.0332  &  0.7649   \\
\end{tabular}
\end{ruledtabular}
\end{table}


\begin{table}
\caption{\label{table_supplN5} Raw data for $N=5$ critical exponents in $d=3$ 
obtained with various regulators at various orders of the DE.
When a value of $\alpha$ different of PMS is employed, this is explicitly indicated.}
\begin{ruledtabular}
\begin{tabular}{lllll}
      & \hspace{-0.5cm}regulator          &  $\nu$         &  $\eta$ &  $\omega$ 
\\
\hline
LPA   &$W$  				  & 0.8395	   &  0      &  0.7706   
 \\
      &$\Theta^1$ \cite{Litim02}          & 0.8377	   &  0      &  0.7667   
 \\
 &$\Theta^3$	                  & 0.8385	   &  0      &  0.7687   
 \\
      &$E$ 	                          & 0.8402	   &  0      &  0.7721	 
\\
\hline   
$O(\partial^2)$
     &$W$ 		       	          & 0.7731         &  0.0420 &  0.7241    \\
     &$\Theta^3$        	 	  & 0.7737	   &  0.0416 &  0.7275    \\
     &$E$          		          & 0.7722	   &  0.0425 &  0.7199	   \\
\hline
$O(\partial^4)$
      &$W$  			          &  0.7808	   & 0.0323  &  0.7584   
 \\
      &$\Theta^3$ 		          &  0.7806	   & 0.0327  &  0.7558   
  \\
      &$E$		                  &  0.7815	   & 0.0313  &  0.7648   
    \\
\end{tabular}
\end{ruledtabular}
\end{table}



\begin{table}
\caption{\label{table_supplN10} Raw data for $N=10$ critical exponents in $d=3$ 
obtained with various regulators at various orders of the DE.
When a value of $\alpha$ different of PMS is employed, this is explicitly indicated.}
\begin{ruledtabular}
\begin{tabular}{lllll}
      & \hspace{-0.5cm}regulator          &  $\nu$         &  $\eta$ &  $\omega$ 
\\
\hline
LPA   &$W$ 				  & 0.9194	   &  0      &  0.8745    \\
      &$\Theta^1$ \cite{Litim02}          & 0.9186	   &  0      &  0.8713    \\
      &$\Theta^3$			  & 0.9190	   &  0      &  0.8729    \\
      &$E$ 	         		  & 0.9198	   &  0      &  0.8758	 \\
      &Power-law \cite{Morris:1997xj}     & 0.94          &  0      &  0.89  \\
      \hline   
$O(\partial^2)$
     &$W$   			          & 0.8774         &  0.0276 &  0.7882    \\
      &$W$ ($\alpha=1$) \cite{VonGersdorff:2000kp} & 0.881     &  0.028 &    \\     
     &$\Theta^3$  	         	  & 0.8775	   &  0.0274 &  0.7903   \\
     &$E$   				  & 0.8772	   &  0.0279 &  0.7853	  \\
     &Power-law \cite{Morris:1997xj}      & 0.95          &  0.0054      &  0.82  \\
     \hline
$O(\partial^4)$
      &$W$			          &  0.8777	   & 0.0222  &  0.8063   \\
      &$\Theta^3$  	   	      &  0.8780	   & 0.0225  &  0.8062     \\
      &$E$	 	              &  0.8771	   & 0.0218  &  0.8081       \\
\end{tabular}
\end{ruledtabular}
\end{table}



\begin{table}
\caption{\label{table_supplN20} Raw data for $N=20$ critical exponents in $d=3$ 
obtained with various regulators at various orders of the DE.
When a value of $\alpha$ different of PMS is employed, this is explicitly indicated.}
\begin{ruledtabular}
\begin{tabular}{lllll}
      & \hspace{-0.5cm}regulator          &  $\nu$         &  $\eta$ &  $\omega$ 
\\
\hline
LPA   &$W$  				  & 0.9610	   &  0      &  0.9384   \\
      &$\Theta^3$		      & 0.9608	   &  0      &  0.9376   \\
      &$E$ 		              & 0.9612	   &  0      &  0.9391	 \\           
      &Power-law \cite{Morris:1997xj}     & 0.96   &  0      &  0.95  \\
      \hline   
$O(\partial^2)$
     &$W$           		  & 0.9414 *     &  0.0149 &  0.8875   \\
     &$\Theta^3$		      & 0.9414	   &  0.0148 &  0.8880    \\
     &$E$      			      & 0.9414 *	   &  0.0151 &  0.8867	   \\
      &Power-law \cite{Morris:1997xj}     & 0.98   &  0.0021      &  0.93  \\
     \hline
$O(\partial^4)$
      &$W$			          &  0.9409	   & 0.0125  &  0.8875    \\
      &$\Theta^3$ 			&  0.9411	   & 0.0126  &  0.8884     \\
      &$E$			    &  0.9406	   & 0.0123  &  0.8863    \\
\end{tabular}
\end{ruledtabular}
\end{table}



\begin{table}
\caption{\label{table_supplN100} Raw data for $N=100$ critical exponents in 
$d=3$ obtained with various regulators at various orders of the DE.
When a value of $\alpha$ different of PMS is employed, this is explicitly indicated.}
\begin{ruledtabular}
\begin{tabular}{lllll}
      & \hspace{-0.5cm}regulator          &  $\nu$         &  $\eta$ &  $\omega$ 
\\
\hline
LPA   &$W$  				  & 0.9925	   &  0      &  0.9882    \\
      &$\Theta^3$		          & 0.9924	   &  0      &  0.9880   \\
      &$E$ 		                  & 0.9925	   &  0      &  0.9883	 \\           
      &Power-law \cite{Morris:1997xj}     & 0.994          &  0      &  0.991  \\
      \hline   
$O(\partial^2)$
     &$W$  				      & 0.98906     &  0.00308 &  0.9781	   \\
     &$W$ ($\alpha=1$) \cite{VonGersdorff:2000kp} & 0.990     &  0.0030 &    \\
     &$\Theta^3$			  & 0.98933	   &  0.00294 &  0.9782 *   \\
     &$E$  			          & 0.98908	   &  0.00310 &  0.9781	   \\
\hline
$O(\partial^4)$
      &$W$			  &  0.98884	   & 0.00263  &  0.9771	   \\
      &$\Theta^3$ 				  &  0.98888	   & 0.00264  &  0.9772     \\
      &$E$			 	      &  0.98877  & 0.00260  &  0.9767      \\
     &Power-law \cite{Morris:1997xj}     & 0.998          &  0.00034      &  0.988  \\
      \end{tabular}
\end{ruledtabular}
\end{table}


\section{Numerical method}\label{Ap:NumParam}

We describe in this section the details of the numerical method used
to determine the fixed points and critical exponents at order
$\mathcal O(\partial^s)$ of the DE approximation within the NPRG with
$s=0$, $s=2$ and $s=4$. The general structure of the procedure can be
split in three steps: 1) deriving the flow equations of each function
in the {\it ansatz} for the effective action; 2) finding the fixed
point which governs the critical behavior of the system and 3)
obtaining the critical exponents from the fixed point solution.

\subsection{Deriving flow equations and truncation}

In order to determine the flow equations for each of the function in
the {\it ansatz} of the effective action Eq.~(\ref{ansatz-order4}), we
compute from this {\it ansatz} the general $n$-point vertex function
$\Gamma_{k\,i_1,\dots,i_n}^{(n)}$ and evaluate it in a homogeneous
field configuration. As a rule of thumb for the DE approximation at
order $\mathcal{O}(\partial^s)$, one needs to compute all $n$-point
vertex functions up to $n=2+s$.  Indeed, this is easy to understand by
noticing that to isolate the flow of all functions, which are
characterized by different internal indices and momentum structures,
one needs to compute the flow of all vertex functions up to
$\Gamma_k^{(s)}$.  However, computing the flow of any $\Gamma_k^{(n)}$
involves the vertex functions $\Gamma_k^{(n+1)}$ and
$\Gamma_k^{(n+2)}$.

We highlight that, when plugging in the vertex functions in the
r.h.s. of the flow equations for the different $\Gamma_k^{(n)}$, we
truncate the product of vertex functions \textit{before expanding
  propagators} at order $s$. This is different from what was usually
done in previous uses of the DE, where all terms coming from the
product were taken into account leading to bigger equations (which are
more complicate to handle). Anyway, although this could be done in
principle, the difference between the two schemes are of order
$\mathcal{O}(p^{s+2})$ which makes the shorter and simpler flow
equations the selected option.

Finally, matching in the l.h.s. and in the r.h.s. of the flow equations the indices and momentum structures. Allows to
compute separately each of the flow equations for the different functions in the {\it ansatz}.

\subsection{Finding the fixed point}

There are two ways to go for finding the fixed point of the flow
equations. The first one, which is more traceable to an experimental
procedure, is to start from a microscopic theory or initial condition
for $\Gamma_{k=\Lambda}$ and integrate the flow equation. One can do
this for different values of the initial conditions and, in
particular, vary or fine-tune one parameter. By a dichotomy procedure
(which can be easily implemented by observing the flow of a certain
quantity, say the derivative with respect to $\rho$ of the potential
at zero field), one can find an initial condition which leads the RG
flow as close to the fixed point as required. This is equivalent to
varying the temperature and measuring the system in order to find the
critical temperature $T_c$. The other method, which is numerically
more efficient, faster and more precise, consists in finding the zeros
of the beta functions. There exist efficient root-finding procedures
which work fine if one initializes the procedure sufficiently close to
the fixed point. We will call this procedure the \textit{root-finding}
algorithm.

Since having a good initial condition from scratch is not simple, we
combined both approaches. The procedure implemented was then to start
with some value of $N$ (say $N=2$) and dimension $d$ (we set from
start $d=3$ and never changed it) and start with a dichotomy
procedure. This only takes a few hours in a personal computer if one
takes a smart {\it ansatz} for the microscopic theory. After few
dichotomies, the algorithm reaches a vicinity of the fixed point and
the root-finding algorithm can be used. Once the fixed point is found,
we use this as an initial condition for the root-finding procedure for
another value of $N$ (say 2.1) (Since the equations are well behaved
for non-integer values, one can take small variations of $N$ and/or
$d$ and trace the fixed point to a new value of interest of $N$ and
$d$.) In our particular case, we varied $N$ and obtained the fixed
point solution for all values of $N$ considered in the article at
$d=3$. Each new value of $N$ is obtained in a few minutes for a given
regulator in a personal computer.

We discretized the $\rho$ variable into a grid of $N_\rho=40$ points
and evolved the flow equations using a fourth order Runge-Kutta with
fixed step with free boundary conditions for the $\rho$ direction.
Because of the procedure used, there was no need to optimize in the
time step taken, this part was merely to find a good enough fixed
point solution for the root-finding part, which was implemented with a
Newton-Raphson algorithm.

The normalization condition is fixed as
$\tilde{Z}(\tilde{\rho}_i)|_{i=N_\rho/4}=1$, where
$\tilde{Z}(\tilde{\rho})$ is the dimensionless version of $Z_k(\rho)$
and $\tilde{\rho}_i$ is the value of $\tilde{\rho}$ at site $i$. On
top of this, the size of the box $L_\rho$ is adjusted for every $N$
value in order for the minimum of the potential to fall always in the
site $i=N_\rho/4$. From this definition, the value of $\eta_k$ was
extracted at every step of the procedure.

In all cases, the momentum integrals were performed using an
adaptative 21 point Gauss-Kronrod quadrature rule (qags) provided in
the quadpack library and $\rho$ derivatives were approximated using a
five point centered discretization except at the borders of the grid
where five points were still used but, of course, not centered for the
first two and last two points in the $\rho$ grid.

\subsection{Obtaining critical exponents}

With a very precise fixed point solution we turn to finding the
critical exponents. As just mentioned, $\eta_k$ is extracted from the
normalization condition and is obtained simultaneously with the fixed
point solution.  Indeed, the factor $Z_k$ is the field renormalization
which is related to the running anomalous dimension by
$\partial_t Z_k=-\eta_k Z_k$ and when approaching the fixed point
$\eta_k$ approaches the field anomalous dimension $\eta$.

For the critical exponents $\nu$ and $\omega$ we performed a linear
stability analysis around the fixed point. We computed the
$\mathcal{M}$ stability matrix by evaluating at perturbed position of
the fixed point and computed the eigenvectors of the $13N_\rho-1$
linear system ($N_\rho$ variables for each function $U$, $Z$, $Y$,
$W_1$,\dots,$W_{10}$). The $-1$ corresponds to the normalization
condition which removes the variable attribute of
$\tilde{Z}(\tilde{\rho}_i)|_{i=N_\rho/4}=1$. The smallest eigenvalue
$\lambda_1$ is identified with $\nu$ as $\lambda_1=-\nu^{-1}$, while
the second smallest eigenvalue is simply $\lambda_2=\omega$.

We also tested that the results that we obtain by diagonalizing the
stability matrix coincide with those corresponding, for example, to
the evolution with $t$ of the derivative of the potential with respect
to $\rho$ at zero field near the fixed point given by
\begin{equation}
 U_k'(\rho=0)\sim U_*'(\rho=0)+A \exp(-t/\nu)+B \exp(t \omega)+\dots
\end{equation}

All our results have been checked against changing parameters in order
to use optimal or near optimal set of parameters. The extent of the
field domain considered was also varied, as well as the accuracy with
which integrals were calculated.

\bibliographystyle{apsrev4-1} 
\bibliography{DE4ON}

\end{document}